\documentclass[10pt]{article}

\usepackage{amsmath, latexsym, amssymb, amscd, psfrag, epsfig}

\numberwithin{equation}{section}
\newcommand{\bi}{\begin{itemize}}
\newcommand{\ei}{\end{itemize}}
\newcommand{\U}{\mathcal{U}}  
\newcommand{\C}{\mathbb{C}}            
\newcommand{\R}{\mathbb{R}}            
\newcommand{\N}{\mathbb{N}}            
\newcommand{\Z}{\mathbb{Z}}           
\newcommand{\Dsum}{\bigoplus}          
\newcommand{\Tens}{\bigotimes}         
\newcommand{\tens}{\otimes}            
\newcommand{\symm}{\mathrm{S}}            
\newcommand{\wlim}{\mathop{\mathrm{w{-}lim}}}

\newcommand{\F}{\mathcal{F}}            
\newcommand{\K}{{\mathcal K}} 
\newcommand{\h}{\mathcal{H}} 

\newcommand{\A}{{\mathcal A}}       

\newcommand{\br}{\left\langle}
\newcommand{\ke}{\right\rangle}

\newcommand{\ra}{\rightarrow}

\newcommand{\om}{\omega}
\newcommand{\omt}{\omega_{\mathbf{t}}}

\newtheorem{theorem}{Theorem}[section]

\newtheorem{definition}[theorem]{Definition}
\newtheorem{lemma}[theorem]{Lemma}

\newtheorem{proposition}[theorem]{Proposition}
\newtheorem{corollary}[theorem]{Corollary}

\def\patrat{{\vcenter{\vbox{\hrule height.4pt \hbox{\vrule width.4pt 
height1.45ex \kern1.45ex \vrule width.4pt}
\hrule height.4pt}}}}

\def\qed{\hfill$\patrat$}

\newcommand{\barint}{\hbox{$\int$\kern-0.75\intwidth\vrule width 0.5\intwidth height 2.4pt depth -2pt\kern0.25\intwidth}}
\newlength\intwidth
\setbox0=\hbox{$\int$}
\intwidth=\wd0

\begin{document}
\title{Generalised Brownian Motion\\
 and Second Quantisation}
\author{\normalsize M\u ad\u alin Gu\c t\u a 
\footnote{E-mail: guta@sci.kun.nl}\and \normalsize Hans Maassen
\footnote{E-mail: maassen@sci.kun.nl}}
\date{\today}
\maketitle

\begin{center}
{\rm Mathematisch Instituut}\\
{\rm Katholieke Universiteit Nijmegen} \\ 
{\rm Toernooiveld 1, 6526 ED Nijmegen}\\
{\rm The Netherlands}\\
{\rm fax}:+31 24 3652140
\end{center}

\begin{abstract}\noindent
A new approach to the generalised Brownian motion introduced by 
M. Bo\.zejko and R. Speicher is described, 
based on symmetry rather than deformation. 
The symmetrisation principle is provided by Joyal's 
notions of tensorial and combinatorial species. 
Any such species $V$ gives rise to an endofunctor $\mathcal{F}_V$ 
of the category of Hilbert spaces with contractions. 
A generalised Brownian motion  is an algebra of creation and 
annihilation operators acting on 
$\mathcal{F}_V(\mathcal{\h})$ for arbitrary Hilbert spaces $\h$ and having
a prescription for the calculation of vacuum expectations 
in terms of a function $\mathbf{t}$ on pair partitions. The positivity
is encoded by a $^*$-semigroup of ``broken pair partitions'' whose 
representation space with respect to $\mathbf{t}$ is $V$. 
The existence of the second quantisation 
as functor $\Gamma_{\mathbf{t}}$ from Hilbert spaces to noncommutative probability spaces is proved to be equivalent to the multiplicative 
property of the function $\mathbf{t}$ . 
For a certain one parameter interpolation between the fermionic and the free 
Brownian motion it is shown that the ``field algebras'' $\Gamma(\K)$ are 
type $\mathrm{II}_1$ factors when $\K$ is infinite dimensional.
\end{abstract}

\section{Introduction}\label{sec.introduction}

In non-commutative probability theory one is interested in finding 
generalisations of classical probabilistic concepts such as 
independence and processes with independent stationary increments. 
Motivated by a central limit theorem 
result and by the analogy with classical Brownian motion, 
M. Bo\.zejko and R. Speicher proposed in \cite{Boz.Sp.1} a 
class of operator algebras called ``generalised Brownian motions'' 
and investigated an example of interpolation between the classical 
\cite{Simon} and the free motion of Voiculescu \cite{Voi.Dy.Ni.}. A better
 known interpolation is provided by the ``$q$-deformed 
commutation relations'' 
\cite{Boz.Ku.Spe., Boz.Sp.2, Boz.Sp.3, Fiv., Fr.Bur., Grb., Maa.vLee., Zag.}. 
Such an operator algebra is obtained by performing the GNS representation 
of the free tensor algebra $\mathcal{A}(\K)$ over an arbitrary 
infinite dimensional real Hilbert space $\K$, with respect to a 
``Gaussian state'' $\tilde{\rho}_{\mathbf{t}}$ defined by the 
following ``pairing prescription'':  
\begin{equation}
\tilde{\rho}_{\mathbf{t}}(\om(f_1)\dots \om(f_n))=
\left\{\begin{array}{ll}
 0 & \textrm{if $n$ odd} \\
\underset{\mathcal{V}\in\mathcal{P}_2(n)}{\sum} \mathbf{t}(\mathcal{V}) 
\underset{(k,l)\in \mathcal{V}}{\prod}\br f_k,f_l\ke
& \textrm{if $n$ even} 
\end{array}\right.
\end{equation}
where $f_i\in\K, \omega(f_i)\in \mathcal{A}(\K)$ and the sum runs over 
all pair partitions of the ordered set $\{1,2,\dots ,n\}$. 
The functional is uniquely determined by the complex valued function 
$\mathbf{t}$ on pair partitions. 
Classical Brownian motion is obtained by
taking $\K=\mathrm{L}^2(\R_+)$ and $B_s:= \om(\mathbf{1}_{[0,s)})$ 
with the constant function $\mathbf{t}(\mathcal{V})=1$ on all pair partitions; 
the free Brownian motion 
\cite{Voi.Dy.Ni.} requires $\mathbf{t}$ to be 0 on crossing partitions 
and 1 on non-crossing partitions.  

If one considers complex Hilbert spaces, the analogue of a Gaussian state is
called a Fock state. We show that the GNS representation of the free
algebra $\mathcal{C}(\h)$ of
creation and annihilation operators with respect 
to a Fock state $\rho_{\mathbf{t}}$ can be described in a functorial way
inspired by the notions of tensorial species of 
Joyal \cite{Joyal, Joyal.2}: the representation space has the form
\begin{equation}
\F_{\mathbf{t}}(\h):=\Dsum_{n=0}^\infty \frac{1}{n!} V_n \tens_s \h^{\tens n}
\end{equation}
where $V_n$ are Hilbert spaces carrying unitary representations of 
the symmetric groups $\mathrm{S}(n)$ and $\tens_s$ means the subspace 
of the tensor product containing vectors which are invariant under the 
double action of $\mathrm{S}(n)$. 
The creation operators have the expression:
\begin{equation}
a^*_{\mathbf{t}}(h) ~v\tens_s (h_0\tens\dots\tens h_{n-1})=
(j_n v)\tens_s (h_0\tens\dots\tens h_{n-1}\tens h_n)
\end{equation}   
where $j_n:V_n\to V_{n+1}$ is an operator which intertwines the action 
of $\mathrm{S}(n)$ and $\mathrm{S}(n+1)$.                          

In Section \ref{sec.semigroup} we connect these Fock representations with 
positive functionals on a certain algebraic object 
$ \mathcal{B}\mathcal{P}_2(\infty)$ which we call the $^*$-semigroup 
of ``broken pair partitions''. 
The elements of this $^*$-semigroup can be described 
graphically as segments located between two vertical lines which cut through 
the graphical representation of a pair partition. In particular, 
the pair partitions are elements of $\mathcal{B}\mathcal{P}_2(\infty)$.
We show that if $\rho_{\mathbf{t}}$ is a Fock state then the function 
$\mathbf{t}$ has a natural extension to a positive functional 
$\hat{\mathbf{t}}$ on $ \mathcal{B}\mathcal{P}_2(\infty)$. 
The GNS-like representation with respect to 
$\hat{\mathbf{t}}$ provides the combinatorial data 
$(V_n, j_n)_{n=0}^\infty$ associated to $\rho_{\mathbf{t}}$.

The representation of $\mathcal{A}(\K)$ with respect to a Gaussian state 
$\tilde{\rho}_{\mathbf{t}}$ is a $^*$-algebra generated by 
``fields'' $\omega_{\mathbf{t}}(f)$. Monomials of such fields can be seen as 
moments, with the corresponding cumulants being a generalisation of the 
Wick products known from the 
$q$-deformed Brownian motion \cite{Boz.Ku.Spe.}. 
Using generalised Wick products we prove that any Gaussian state 
$\tilde{\rho}_{\mathbf{t}}$ extends to a Fock state 
$\rho_{\mathbf{t}}$ on the 
algebra of creation and annihilation operators $\mathcal{C}(\K_\C)$
(see section \ref{gen.wick.products}). 

Second quantisation is a special type of \textit{functor of white noise}, a 
functor from the category of real Hilbert spaces with contractions to
the category of (non-commutative) probability spaces. The underlying idea is 
to use the field operators $\omt (\cdot)$ to construct 
von Neumann algebras $\Gamma_{\mathbf{t}}(\K)$ for any real Hilbert space
$\K$ and a fixed positive definite functions $\mathbf{t}$. The question is 
for which $\mathbf{t}$ one can carry out 
the construction of such a functor $\Gamma_{\mathbf{t}}$. 
From general considerations on functors of second quantisation we obtain that
the function $\mathbf{t}$ must have the multiplicative property, 
a form of statistical independence. Conversely, for multiplicative 
$\mathbf{t}$ the field operators are essentially selfadjoint, and provide a 
natural definition of the von Neumann algebra $\Gamma_{\mathbf{t}}(\K)$ .  
The second step is the implemetation of the second quantisation 
$\Gamma_{\mathbf{t}}(T)$ of an arbitrary contraction $T$ between Hilbert 
spaces. This is done separately for isometries and coisometries which are
then used to define the second quatisation for arbitrary contractions.

In the last section we develop a useful criterion, 
in terms of the spectrum of a characteristic contraction, 
for factoriality of the algebras $\Gamma_\mathbf{t}(\ell^2(\Z))$ 
in the case when the vacuum state $\rho_{\mathbf{t}}$ is tracial. 
We then apply it to a particular example of positive definite function 
$\mathbf{t}_q$ where $0\leq q<1$, which interpolates between the 
bosonic and free cases and has been introduced in \cite{Boz.Sp.1} 
(see \cite{Gu.Maa.} for another proof of the positivity). 
We conclude that $\Gamma_\mathbf{t}(\ell^2(\Z))$ is a type 
$\mathrm{II}_1$ factor. 
Further generalisation of this criterion to factors of type $\mathrm{III}$ 
will be investigated in a forthcoming paper \cite{Gu.Maa.2}.

\section{Definitions and description of the Fock representation}
\label{SecRepresentations}

The generalised Brownian motions \cite{Boz.Sp.1} are 
representations with respect to special \textit{gaussian} states 
on free algebras over real Hilbert spaces. 
We start by giving all necessary definitions and subsequently 
we will analyse the structure of the \textit{Fock representations} which 
are intimately connected with the generalised Brownian motion 
(see section \ref{gen.wick.products}) .  
\begin{definition}{\rm
Let $\K$ be a real Hilbert space. The algebra $\mathcal{A}(\K)$ 
is the free unital $^*$-algebra with generators $\om (h)$ for all
 $h \in \K$, divided by the relations:
\begin{equation}
\om(af+bg)=a\om (f)+ b\om (g),   \qquad   \om (f)=\om (f)^*
\end{equation}
for all $f,g\in\K$ and $a,b\in\R$.}
\end{definition}

\begin{definition}{\rm
Let $\h$ be a complex Hilbert space. The algebra $\mathcal{C}(\h)$
 is the free unital $^*$-algebra with generators $a(h)$ and $a^*(h)$ 
for all $h \in \h$, divided by the relations:
\begin{equation}
a^*(\lambda f+\mu g)=\lambda a^*(f)+ \mu a^*(g),\qquad a^*(f)= a(f)^*
\end{equation}
for all $f,g\in\h$ and $\lambda, \mu\in\C$.}
\end{definition}

\noindent 
We notice the existence of the canonical injection from 
$\mathcal{A}(\K)$ to $\mathcal{C}(\K_\C)$
\begin{equation}
\om (h)\mapsto a(h)+ a^*(h)
\end{equation}
where $\K_\C$ is the complexification of the real Hilbert space $\K$. On the 
algebras defined above we would like to define positive linear functionals 
by certain pairing prescriptions for which we need some notions 
of pair partitions.

\begin{definition}{\rm
Let $S$ be a finite ordered set. We denote by $\mathcal{P}_2(S)$ 
is the set of pair partitions of $S$, that is 
$\mathcal{V}\in\mathcal{P}_2(S)$ if $\mathcal{V}$ consists of $\frac{1}{2}n$
disjoint ordered pairs $(l,r)$ with $l<r$ having $S$ as their reunion. 
The set of all pair partitions is}
\begin{equation}
\mathcal{P}_2(\infty):=\bigcup_{r=0}^\infty\mathcal{P}_2 (2r).
\end{equation} 
\end{definition}    
Note that $ \mathcal{P}_2(n)=\emptyset$ if $n$ is odd. 
In this paper the symbol $\mathbf{t}$ will always stand for 
a function $\mathbf{t}:\mathcal{P}_2(\infty)\ra \C$. We will always choose 
the normalisation $\mathbf{t}(p)=1$ for $p$ the pair partition containing 
only one pair.

\begin{definition}\label{def.Fock.state}{\rm
A \textit{Fock state} on the algebra $\mathcal{C}(\h)$ is a
 positive normalised linear functional 
$\rho_{\mathbf{t}}:\mathcal{C}(\h)\to \C$ of the form
\begin{equation}\label{fockstate}
\rho_{\mathbf{t}}(a^{\sharp_1}(f_1)\dots a^{\sharp_n}(f_n))=
\underset{\mathcal{V}\in\mathcal{P}_2(n)}{\sum}\mathbf{t}(\mathcal{V})
\underset{(k,l)\in\mathcal{V}}{\prod}\br f_k,f_l \ke \cdot Q(\sharp_k,\sharp_l)
\end{equation}
the symbols $\sharp_i$ standing for creation 
or annihilation and the two by two covariance matrix $Q$ is given by
\begin{displaymath}
Q=
\left(\begin{array}{ccc}
\rho(a_ia_i) & \rho(a_ia^*_i) \\
\rho(a^*_ia_i)  &  \rho(a^*_ia^*_i)
\end{array}\right)
= 
\left(\begin{array}{ccc}
 0 & 1 \\
 0 & 0
\end{array}\right).
\end{displaymath}
where $a_i=a(e_i)$ and $e_i$ is an arbitrary 
normalized vector in $\h$. Note that the l.h.s. of (\ref{fockstate}) is zero for odd values of $n$.} 
\end{definition}

\begin{definition}\label{def.Gaussian.state}{\rm
A \textit{Gaussian state} on $\mathcal{A}(\K)$ is a 
positive normalised linear functional $\tilde{\rho}_{\mathbf{t}}$ with moments

\begin{equation}\label{gaussianstate}
\tilde{\rho}_{\mathbf{t}}(\om(f_1)\dots \om(f_n))=
\underset{\mathcal{V}\in\mathcal{P}_2(n)}{\sum} \mathbf{t}(\mathcal{V}) 
\underset{(k,l)\in \mathcal{V}}{\prod}\br f_k,f_l\ke
\end{equation}
\noindent }
\end{definition}

\noindent
\textbf{Remark.} The restriction of a Fock state $\rho_{\mathbf{t}}$ on
 $\mathcal{C}(\K_\C)$ to the subalgebra $\mathcal{A}(\K)$ is 
the Gaussian state $\tilde{\rho}_{\mathbf{t}}$. If $\rho_{\mathbf{t}}$ 
is a Fock state for all choices of $\K$ then we call the function
\begin{displaymath}
\mathbf{t}:\mathcal{P}_2(\infty)\ra \C 
\end{displaymath}
\textit{positive definite}.

\noindent
The GNS representations associated to pairs
 $(\mathcal{C}(\h), \rho_{\mathbf{t}})$ have been studied
 in a number of cases. One obtains a representation $\pi_{\mathbf{t}}$ of
 $\mathcal{C}(\h)$ as  $^*$-algebra of creation and annihilation
 operators acting on a Hilbert space $\F_{\mathbf{t}}(\h)$ which has a
 Fock-type structure
\begin{displaymath}
\F_{\mathbf{t}}(\h)=\Dsum_{n=0}^\infty \h_n
\end{displaymath} 
with $\h_n$ being a (symmetric) subspace of $\h^{\tens n}$ in
 the case of bosonic or fermionic algebras \cite{Simon}, the full tensor 
product in models of free probablity \cite{Voi.Dy.Ni.}, 
a deformation of it in the case of q-deformations 
\cite{Boz.Ku.Spe., Boz.Sp.2, Boz.Sp.3, Fiv., Fr.Bur., Grb., Maa.vLee., Zag.}, 
or even ``larger'' spaces containing more copies of $\h^{\tens n}$ 
with a deformed inner product in the case of another deformation 
depending on a parameter $-1\leq q\leq 1$ constructed in \cite{Boz.Sp.1}. 
The action of the creation operators is $a^*(f)\Omega_{\mathbf{t}}=f\in \h,$
\begin{displaymath}
a^*(f)f_1\tens\dots\tens f_n=f\tens f_1\tens\dots\tens f_n
\end{displaymath}
while that of the annihilation operator is less transparent, 
depending on the inner product on $\h_n$. Proving the positivity of this 
inner product is in general nontrivial.

\vspace{2mm}

In \cite{Gu.Maa.} we have followed a different, more combinatorial 
approach to the study of the representations 
$\pi_{\mathbf{t}}(\mathcal{C}(\h))$ for various examples of 
positive definite functions $\mathbf{t}$. 
We give here a brief description of our construction. 
The representation space is denoted by $\F_V(\h)$ 
and has certain symmetry properties encoded by a sequence 
$(V_n)_{n=0}^\infty$ of (not necessarily finite dimensional) 
Hilbert spaces such that each $V_n$ carries a unitary representation of 
the symmetric group $\symm (n)$
\begin{equation}
\symm (n)\ni \pi\mapsto U(\pi)\in \mathcal{U}(V_n).
\end{equation}

\noindent
In concrete examples we have realised $V_n$ as $\ell^2(F[n])$ 
where  $F[~]$ is a \textit{species of structures} 
\cite{Ber.Lab.Ler., Joyal, Joyal.2}, i.e., 
a functor from the category of finite sets with bijections as morphisms to 
the category of finite sets with maps as morphisms. For each finite set 
$A$, the rule $F$ prescribes a finite set $F[A]$ whose elemens are called 
$F$-structures over the set $A$. Moreover for any bijection $\sigma:A\to B$
 there is a map $F[\sigma]:F[A]\to F[B]$ such that 
$F[\sigma\circ\tau]= F[\sigma]\circ F[\tau]$ and 
$F[\mathrm{id}_A]=\mathrm{id}_{F[A]}$. In particular for 
$n:=\{0,1,\dots n-1\}$ there is an action of the symmetric group 
$\symm (n)$ on the set of structures:
\begin{displaymath}
\forall \pi\in\symm (n),\qquad F[\pi]:F[n]\ra F[n]
\end{displaymath}
which gives a unitary representation $U(\cdot)$ of $\symm (n)$ on $V_n:=\ell^2(F[n])$. Simple examples are such species as sets, ordered sequences, trees, graphs, etc. 

\noindent 
We define 
\begin{equation}\label{defsymm}
\F_V(\h):=\Dsum_{n=0}^\infty \frac{1}{n!} V_n \tens_s \h^{\tens n}
\end{equation}
where $V_n \tens_s \h^{\tens n}$ is the subspace of 
$V_n \tens \h^{\tens n}$ spanned by the vectors $\psi$ 
invariant under the  action of $\symm (n)$:
\begin{displaymath}
\psi=(U(\pi)\tens\tilde{U}(\pi))\psi, \qquad\mathrm{for~ all}~\pi\in\symm (n)
\end{displaymath}
with $\tilde{U}(\pi)\in\mathcal{U}(\h^{\tens n})$,  
\begin{equation} 
\tilde{U}(\pi): h_0\tens\dots \tens h_{n-1}  \mapsto  h_{\pi^{-1}(0)}
\tens\dots \tens h_{\pi^{-1}(n-1)},
\end{equation}
the factor $\frac{1}{n!}$ referring to the inner product. 
The symmetric Hilbert space $\F_V(\h)$ is spanned by linear
 combinations of vectors of the form:
\begin{equation}
v\tens_s h_0\tens\dots\tens h_{n-1}:=
\frac{1}{n!}\sum_{\pi\in \symm (n)}U(\pi)v
\tens\tilde{U}(\pi)h_0\tens\dots\tens h_{n-1}.
\end{equation} 
The creation and annihilation operators are defined with the
 help of a sequence of densely defined linear maps $(j_n)_{n=0}^\infty$ with
$j_n:V_n\ra V_{n+1}$ satisfying the intertwining relations
\begin{equation} \label{intertwining}
j_n \cdot U(\pi)=U(\iota_n(\pi))\cdot j_n, \qquad\forall \pi\in \symm (n)
\end{equation}
with $\iota_n:\symm (n)\ra \symm (n+1)$ being the canonical 
embedding associated to the inclusion of sets
\begin{equation}
n :=\{0,1,\dots ,n-1\}\hookrightarrow n+1 :=\{0,1, \dots, n\}. 
\end{equation}
In the examples using species of structures the map 
$j_n:\ell^2(F[n])\to\ell^2(F[n+1])$ 
is constructed by giving the matrix elements 
$j_n(s,t):=\br\delta_t,V_n \delta_s\ke$ 
which can be seen as ``transition coefficients'' between $s\in F[n]$ and 
$t\in F[n+1]$. For example \cite{Gu.Maa.} if the species $F[\cdot]$ 
is that of rooted trees 
one can choose $j_n(s,t)=1$ if the tree $s$ is obtained by removing the leaf
 with label $n$ from the tree $t$; otherwise we choose $j_n(s,t)=0$. 
Notice that there is no canonical manner of defining $j_n$ but certain species
of structures offer rather natural definitions, 
for example the species of sets, ordered sequences, rooted trees, 
oriented graphs, 
sequences of non empty sets, etc \cite{Gu.Maa.}. \\
Let $h\in\h $; 
the creation operator $a^*_{V,j}(h)$ has the action: 
\begin{equation}\label{defcr}
a^*_{V,j}(h) v\tens_s (h_0\tens\dots\tens h_{n-1}):=
 (j_n v) \tens_s (h_0\tens\dots\tens h_{n-1}\tens h).
\end{equation} 
The annihilation operator $a_{V,j}(h)$ is the adjoint of $a^*_{V,j}(h)$. 
Its action on the $n+1$-th level is given by the restriction 
of the operator 
\begin{eqnarray}\label{defann}
\tilde{a}_{V,j}(h):
V_{n+1}\tens \h^{\tens n+1}& \to& 
V_{n}\tens \h^{\tens n}             \nonumber\\
                   v\tens (h_0\tens\dots\tens h_n)& \mapsto& 
\br h,h_n \ke j^*_nv\tens (h_0\tens\dots\tens h_{n-1}).
\end{eqnarray} 
to the subspace $V_{n+1}\tens_s\h^{\tens n+1}$. 
Note that due to condition (\ref{intertwining}) the operators 
$a_{V,j}^*(h),a_{V,j}(h)$ are well defined. 
Let us denote by $\mathcal{C}_{V,j}(\h)$ the $^*$-algebra generated by
 all operators $a^*_{V,j}(h), a_{V,j}(h)$ and by $\Omega_V\in V_0$ the 
normalised vacuum vector in $\F_V(\h)$. The following theorem is a 
generalisation of Proposition 5.1 in \cite{Gu.Maa.}:

\begin{theorem}\label{prop.combinatorics}
Let $(\F_V(\h),\mathcal{C}_{V,j}(\h), \Omega_V )$ be a representation of 
$\mathcal{C}(\h)$ as described above, then the state 
$\rho_{V,j}(\cdot)=\br\Omega_V,\cdot \Omega_V\ke$ is a Fock state, i.e. 
there exists a positive definite function $\mathbf{t}$ on pair partitions
depending on $(V_n,j_n)_{n=0}^\infty$ such that $\rho_{V,j}=\rho_{\mathbf{t}}$.
\end{theorem}

\noindent
\textit{Sketch of the proof.} Let $A\in\mathcal{B}(\h)$. On $\F_V(\h)$ 
we define the operator
\begin{equation}
\mathrm{d}\Gamma_V(A): v_n\tens_{s} f_0\tens\dots \tens f_{n-1}\mapsto
\sum_{k=0}^{n-1} v_n\tens_{s} f_0\tens\dots \tens Af_k \tens\dots\tens f_{n-1}
\end{equation}
for $v_n\in V_n, f_i\in\h$. Then the following commutation relations hold:
\begin{equation}\label{comm.rel1}
 ~[a_{V,j}(f), \mathrm{d}\Gamma_V(A)]  =  a_{V,j}(A^*f).       
\end{equation}
In particular by choosing an orthonormal basis 
$\{e_i\}_{i\in I} $ in $\h$ and denoting 
$a_i^\sharp :=a^\sharp_{V,j}(e_i)$ we obtain for all $i_k\neq i_0$
\begin{equation}\label{eq.comm}
[\mathrm{d}\Gamma_V(|e_{i_0} \rangle \langle e_i|), 
a^{\sharp_k}_{i_k}]=
\delta_{i_k,i}\cdot\delta_{\sharp_k, *} \cdot a^*_{i_0}.
\end{equation}
Let $\psi=\left(\prod_{k=1}^{n}a^{\sharp_k}_{i_k}\right)\Omega_V$. Then 
$a_{i_0}\psi =0$ if $i_0\neq i_k$ for all $k =1,\dots, n$. By using 
(\ref{comm.rel1}), it follows that 
\begin{equation}
a_i\psi=
[a_{i_0}, \mathrm{d}\Gamma (|e_{i_0} \rangle \langle e_i|)]\psi =
a_{i_0} \mathrm{d}\Gamma(|e_{i_0}\rangle \langle e_i|)\psi.   
\end{equation}
We then apply  (\ref{eq.comm}) repatedly to obtain
\begin{equation}\label{eq.derivation}
a_i\left(\prod_{k=1}^{n}a_{i_k}^{\sharp_k}\right)\Omega_V = \sum_{k=1}^{n}
\delta_{i,i_k}\cdot\delta_{\sharp_k, *}\cdot a_{i_0}
\left(\prod_{p=1}^{k-1} a_{i_p}^{\sharp_p}\right) 
\cdot a_{i_0}^{*}\cdot
\left(\prod_{q=k+1}^{n} a_{i_q}^{\sharp_q}\right)\Omega_V . 
\end{equation}
The vacuum expectation of a monomial $\prod_{k=1}^{n}a_{i_k}^{\sharp_k}$ 
can be different from zero only if the number of creators is 
equal to the number of annihilators, $a_{i_1}^{\sharp_1}$ is an 
annihilator and  $a_{i_n}^{\sharp_n}$ a creator. We will 
therefore assume that this is the case. 
We put the monomial in the form $a_{i_1}\prod_{k=2}^{n}a_{i_k}^{\sharp_k}$
 and  apply (\ref{eq.derivation}). We obtain a sum over all pairs 
$(a_{i_1}, a_{i_k}^*)$ of the same color ($i_1=i_k$) and replace $i_1$ by
 a new color $i_0$. We pass now to 
the next annihilator in each term of the sum and repeat the procedure, 
the new color which we add this time being different from all the 
colors used previously. After $\frac{n}{2}$ steps we obtain a sum
 containing all possible pairings of annihilators and creators
 of the same color in $ \prod_{k=1}^{n}a_{i_k}^{\sharp_k}$:
\begin{equation}
\rho_{V,j}(\prod_{k=1}^{n}a_{i_k}^{\sharp_k})= 
\sum_{\mathcal{V}\in\mathcal{P}_2(n)}~\prod_{k,l\in\mathcal{V}}
\delta_{i_{k},i_{l}}\cdot 
Q(\sharp_{k},\sharp_{l})\cdot t(\mathcal{V}) 
\end{equation}  
with $t(\mathcal{V}):=\rho_{V,j}(\prod_{k=1}^{n}a_{j_k}^{\sharp_k})$,
where the indices $j_k, \sharp_k$ satisfy the following conditions: 
if $k\neq l$ then $j_{k}=j_{l}$ if and only if $(k,l)\in \mathcal{V}$,
in which case $a_{j_k}^{\sharp_{k}}$ is annihilator 
and $a_{j_l}^{\sharp_{l}}$ is creator.

\qed

\noindent
We prove now that the converse is also true.

\begin{theorem}\label{th.GNSrep}
Let $\mathbf{t}$ be a positive definite function on pair partitions. 
Then for any complex Hilbert space $\h$ the GNS-representation of 
$(\mathcal{C}(\h), \rho_{\mathbf{t}})$ is unitarily equivalent to 
$(\F_V(\h),\mathcal{C}_{V,j}(\h), \Omega_V )$ for a 
sequence $(V_n,j_n)_{n=0}^\infty$ 
dependent only up to unitary equivalence on $\mathbf{t}$.  
\end{theorem}

\noindent
\textit{Proof.} We first consider $\h:=\ell^2(\N^*)$ with the orthonormal basis 
$(e_i)_{i=1}^\infty$. We split the proof in 3 steps.

\noindent
1. Identify the spaces $V_n$ and the maps $j_n$.

\noindent 
Let $(\F_{\mathbf{t}}(\h),\mathcal{C}_{\mathbf{t}}(\h), \Omega_{\mathbf{t}} )$ be the triple obtained 
from the GNS-construction. 
Let $V_n$ be the closure of the subspace of 
$\F_{\mathbf{t}}(\h)$ spanned by vectors
 of the form $v_n:=(\prod_{k=1}^{2p+n}a^{\sharp_k}_{\mathbf{t}}(e_{i_k}))\Omega_{\mathbf{t}}$ 
for which the following conditions hold:

(i) in the sequence $(a^{\sharp_k}_{\mathbf{t}}(e_{i_k}))_{k=1}^{2p+n}$ each 
creation operator $a^*_{\mathbf{t}}(e_j)$ appears exactly once for 
$1\leq j \leq n$; 

(ii) the rest of the sequence contains $p$ creation operators
 $(a^*_{\mathbf{t}}(e_{l_q}))_{q=1}^p$ and  $p$ annihilation operators 
$(a_{\mathbf{t}}(e_{l_q}))_{q=1}^p$ for $p$ vectors $(e_{l_q})_{q=1}^p$ 
different among each other and with $l_q\notin \{1,\dots ,n\}$ 
for all $1\leq q\leq p$. The vector $v_n$ does not depend in fact 
on the colours $(l_q)_{q=1}^p$ but only on the positions of the 
creation and annihilation operators in the monomial. Thus when necessary we 
can consider $l_q>N$ for all $1\leq q \leq n $ and some fixed big enough 
$N\in \N$ . \\
The map $j_n$ is defined as the restriction of $a^*_{\mathbf{t}}(e_{n+1})$ to $V_n$:
\begin{displaymath}
j_n\prod_{k=1}^{2p+n}a^{\sharp_k}_{\mathbf{t}}(e_{i_k})\Omega_{\mathbf{t}}=
a^*_{\mathbf{t}}(e_{n+1})\prod_{k=1}^{2p+n}a^{\sharp_k}_{\mathbf{t}}(e_{i_k})\Omega_{\mathbf{t}}.
\end{displaymath}  
Obviously, the image of $j_n$ lies in  $V_{n+1}$. 

\noindent
The state $\rho_{\mathbf{t}}$ is invariant under unitary transformations 
$U\in\mathcal{U}(\h)$:
\begin{displaymath}
\rho_{\mathbf{t}}(\prod_{k=1}^n a^{\sharp_k}(e_{i_k}))
=\rho_{\mathbf{t}}(\prod_{k=1}^n a^{\sharp_k}(Ue_{i_k})).
\end{displaymath}
Thus 
\begin{equation}\label{def.secquant.Hilbert}
\F_{\mathbf{t}}(U):\prod_{k=1}^n a^{\sharp_k}_{\mathbf{t}}(e_{i_k})\Omega_{\mathbf{t}}\mapsto 
\prod_{k=1}^n a^{\sharp_k}_{\mathbf{t}}(Ue_{i_k})\Omega_{\mathbf{t}}
\end{equation}
is unitary and $\F_{\mathbf{t}}(U_1)\F_{\mathbf{t}}(U_2)=\F_{\mathbf{t}}(U_1U_2)$ for two unitaries 
$U_1, U_2$. The action on the algebra of creation and annihilation operators is
\begin{equation}\label{defF_t}
\F_{\mathbf{t}}(U)a^\sharp_{\mathbf{t}}(f)\F_{\mathbf{t}}(U^*)=a^\sharp_{\mathbf{t}}(Uf).
\end{equation}

\noindent
Considering unitaries which act by permuting the basis vectors 
$\{e_1,\dots ,e_n\}$ and leave all the others invariant
 we obtain a unitary representation of 
$\symm (n)$ on $V_n$. The intertwining property (\ref{intertwining}) follows 
immediately from the definition of $j_n$. Having the ``combinatorial data''
$(V_n,j_n)$, we can construct the triple 
$(\F_V(\h),\mathcal{C}_{V,j}(\h), \Omega_V )$ according to equations 
(\ref{defsymm}, \ref{defcr}, \ref{defann}). 
Similarly to $\F_{\mathbf{t}}(U)$ we have the unitary
\begin{eqnarray}\label{defsecquant}
\F_{V}(U) :\F_{V}(\h) & \ra & \F_{V}(\h)\nonumber\\
           v\tens_s (h_0\tens\dots h_{n-1}) &\mapsto & 
           v\tens_s (U h_0\tens\dots\tens U h_{n-1}) 
\end{eqnarray}
for $U\in\U(\h), v\in V_n$. We call $F_{V}(U)$ the 
\textit{second quantisation} 
of $U$ at the Hilbert space level. Its action on operators is:
\begin{equation}\label{defF_V}
\F_V(U)a^\sharp_{V,j}(f)\F_V(U^*)=a^\sharp_{V,j}(Uf).
\end{equation}

\noindent
Analogously to $V_n$ we define for any finite subset 
$\{i_1,\dots ,i_n\}\subset \N$ the linear subspace 
$V(i_1,\dots ,i_n)$ of $\F_{\mathbf{t}}(\h)$ spanned 
by applying to the vacuum $\Omega_{\mathbf{t}}$ monomials 
$\prod_{k=1}^{2p+n} a^{\sharp_k}_{\mathbf{t}}(e_{j_k})$ for 
which the colours $(j_k)_{k=1}^{2p+n}$ satisfy conditions similar 
to i), ii) but now with $\{i_1,\dots ,i_n\}$ instead of $\{1,\dots ,n\}$. 
For a unitary $U$ which permutes the basis vectors, $Ue_i=e_{u(i)}$ we get
\begin{equation}
\F_{\mathbf{t}}(U)V(i_1,\dots ,i_n)=V(u(i_1),\dots ,u(i_n)).
\end{equation}
One can check by calculating inner products that any two such spaces 
are either orthogonal or coincide. 
Similarly, we define the following subspaces of $\F_{V}(\h)$
\begin{equation}
\tilde{V}(i_1,\dots ,i_n):=
\overline{\mathrm{lin}\{v\tens_s(e_{i_1}\tens\dots\tens e_{i_n}):
\qquad v\in V_n\}}
\end{equation}
\noindent
which are also orthogonal for different sets of ``colours'' 
$\{i_1,\dots ,i_n\}$.\\
2. We proceed by proving 
the equality of the states $\rho_{\mathbf{t}}$ and $\rho_{V,j}$.

\noindent
As $\rho_{V,j}$ is a Fock state by Theorem \ref{prop.combinatorics}, 
we need only verify that the positive definite function 
$\mathbf{t}$ we have started with and 
the one derived from $\rho_{V,j}$ coincide. 
By definition there is an isometry
\begin{eqnarray}
T_n:
V_n & \ra & \F_{V,j}(\h)\nonumber\\
  v & \mapsto & v\tens_s (e_1\tens\dots\tens e_n).
\end{eqnarray}
Furthermore for any unitary $U\in\U(\h)$ which permutes 
the basis vectors such that $Ue_k=e_{i_k} $, the operator
\begin{displaymath}
T(i_1,\dots ,i_n ):
V(i_1,\dots ,i_n ) \ra \tilde{V}(i_1,\dots ,i_n )
\end{displaymath}
\noindent
defined by
\begin{equation}\label{defT}
T(i_1,\dots ,i_n ):=\F_{V}(U)T_n \F_{\mathbf{t}}(U^*)
\end{equation}
depends only on the set $\{i_1,\dots ,i_n\}$. Finally, 
the definitions of $j_n, a^\sharp_{V,j}(f)$ amounts to the 
fact that the following diagram commutes 
\begin{equation}
\begin{CD}
V_n   @>T_n>>   \tilde{V}_n\\ 
@V{a^*_{\mathbf{t}}(e_{n+1})}VV                  @VV{a^*_{V,j}(e_{n+1})}V\\ 
V_{n+1}               @>T_{n+1}>>   \tilde{V}_{n+1} \\
\end{CD}
\end{equation}

\noindent
and by acting from the left and from the right with the appropriate 
second quantisation operators and using 
(\ref{defT}, \ref{defF_t}, \ref{defF_V}) we obtain
\begin{equation}
\begin{CD}
V(i_1,\dots ,i_n)         
@>T(i_1,\dots ,i_n)>>   
\tilde{V}(i_1,\dots ,i_n)\\ 
@V{a^*_{\mathbf{t}}(e_{i_{n+1}})}VV            
@VV{a^*_{V,j}(e_{i_{n+1}})}V\\ 
V (i_1,\dots ,i_{n+1}) 
@>T(i_1,\dots ,i_{n+1})>>
\tilde{V}(i_1,\dots ,i_{n+1})\\
\end{CD}
\end{equation}
with a similar diagram for the annihilation operators. 
This is sufficient for proving the equality
$\rho_{\mathbf{t}}(\prod_{k=1}^{2n} a^{\sharp_k}_{\mathbf{t}}(e_{i_k}))=
\rho_{V,j}(\prod_{k=1}^{2n} a^{\sharp_k}_{V,j}(e_{i_k}))$ 
for monomials containing $n$ pairs of creation and annihilation 
operators of $n$ different colours.\\
3. Finally we prove that $\Omega_{V,j}$ is cyclic vector for 
$\mathcal{C}_{V,j}(\h)$. 

\noindent
The space $\F_V(\h)$ has a decomposition with respect to occupation numbers
\begin{displaymath}
\F_V(\h)=\Dsum_{\{n_1,\dots ,n_k\}}\F_V(n_1, \dots ,n_k)
\end{displaymath}
with 
\begin{equation}\label{eq.occunumb}
\F_V(n_1, \dots ,n_k)=
\overline{\mathrm{lin}\{v\tens_s(\underbrace{e_1\tens\dots\tens e_1}_{n_1}
\tens\dots\tens\underbrace{e_k\tens\dots\tens e_k}_{n_k}, 
v\in V_{n_1+\dots +n_k}\}}.
\end{equation}
We recall that $\tilde{V_n}=\F_V(\underbrace{1,\dots ,1}_{n})$ is spanned 
by linear combinations of vectors of the form  
\begin{displaymath}
\prod_{k=1}^{2p+n}a^{\sharp_k}_{V,j}(e_{i_k})\Omega_{V}=
v\tens_s(e_1\tens\dots\tens e_n)
\end{displaymath}
with monomials satisfying the conditions i) and ii). 
By replacing the creation operators $(a^*(e_k))_{k=1}^n$ 
appearing in the monomial, with the sequence containing 
$n_i$ times the creator $a^*(e_i$) for $i\in\{1,\dots p\}$ and 
$\sum_{i=1}^{p} n_i=n$ we obtain a set of vectors which are dense in 
$\F_V(n_1,\dots ,n_p)$ and this completes the proof of the cyclicity of 
the vacuum.
Putting together 1., 2. and 3. we conclude that the representations 
 $(\F_{\mathbf{t}}(\h),\mathcal{C}_{\mathbf{t}}(\h), \Omega_{\mathbf{t}} )$ 
and 
$(\F_V(\h),\mathcal{C}_{V,j}(\h), \Omega_V )$  are unitarily equivalent for 
infinite dimensional $\h$. The case $\h$ finite dimensional follows by 
restriction of the previous representations to the appropriate subspaces. 

\qed

\section{The $^*$-semigroup of broken pair partitions}
\label{sec.semigroup}

The content of the last two theorems can be summarised 
by the following fact: 
there exist a bijective correspondence between 
positive definite functions on pair partitions $\mathbf{t}$, and 
``combinatorial data'' $(V_n, j_n)_{n=0}^\infty$. 
This suggests that the positivity of $\mathbf{t}$ can be characterised 
in a simpler way by regarding $\mathbf{t}$ as a positive functional on an 
algebraic object containing $\mathcal{P}_2(\infty)$ as a subset. 
Theorem 1 of \cite{Boz.Sp.1} shows that 
a positive definite function on pair partitions $\mathbf{t}$ 
restricts to positive 
definite functions on the symmetric groups $\symm (n)$ for all $n\in\N$ 
through the embedding
\begin{equation}
\symm (n)\ni\tau\mapsto \mathcal{V}_\tau\in\mathcal{P}_2(n)
\end{equation}
given by
\begin{equation}
\mathcal{V}_\tau:=\{(i,2n+1-\tau(i)):~i=1,\dots ,2n\}.
\end{equation}
However $\mathbf{t}$ is not determined completely by its restriction 
and thus one would like to find another algebraic object which completely
encodes the positivity requirement.    
We will show that this is the $^*$-semigroup of \textit{broken pair partitions}
which we denote by $\mathcal{B}\mathcal{P}_2(\infty)$ and will be described
below. Pictorially, the elements  of the semigroup are segments 
obtained by sectioning pair partitions with vertical lines.

\begin{definition}\label{def.semigroup}
Let $X$ be an arbitrary finite ordered set and $(L,P,R)$ a disjoint partition 
of $X$. We consider all the triples $(\mathcal{V}, f_l,f_r)$ where
$\mathcal{V}\in\mathcal{P}_2(P)$ and 
\begin{equation}
f_l:L\rightarrow\{1,\dots , |L|\}, \qquad
f_r:R\rightarrow\{1,\dots , |R|\}
\end{equation} 
are bijections. Any order 
preserving bijection $\alpha :X\to Y$ induces an obvious map   
\begin{equation}
(\mathcal{V}, f_{l},f_{r})\to 
(\alpha\circ\mathcal{V}, f_{l}\circ \alpha^{-1}, f_{r}\circ \alpha^{-1})  
\end{equation}
where $\alpha\circ\mathcal{V}:=\{(\alpha(a),\alpha(b)): 
(a,b)\in\mathcal{V}\}$. 
This defines an equivalence relation; an element $d$ of 
$\mathcal{B}\mathcal{P}_2(\infty)$ is an equivalence class of triples 
$(\mathcal{V}, f_l,f_r)$ under this equivalence relation. 
\end{definition}
 
\begin{figure}[h]
\begin{center}
\includegraphics[width=4cm]{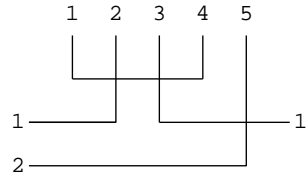}
\end{center}
\caption{Diagram corresponding to an element of the semigroup}
\label{diagram}
\end{figure}

\noindent  
We have the following pictorial representation: an element $d$ is given by 
a diagram containing a sequence of $l+r+2n$
points displayed horizontally with $2n$ of them connected into $n$ pairs, 
$l$ points are connected with other $l$ points 
vertically ordered on the left side (left legs) and $r$ points are 
connected with $r$ points vertically ordered on the right (right legs). 
An example is given in Figure \ref{diagram}. 
In this case we have $X=\{1,\dots,5\}$, $\mathcal{V}=\{(1,4)\}$, 
the left legs are connecting the points labeled 2 and 5 on the horizontal to 
the the points on the left side which are ordered vertically and 
labeled by 1 and 2. Similarly for the right legs. Usually we will 
label the ordered set of horizontal points will be of the form 
$\{n, n+1,\dots n+m\}$.  

\begin{figure}[h]
\begin{center}
\includegraphics[width=10cm]{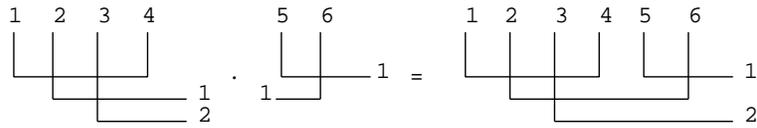}
\end{center}
\caption{Product of two elements}
\label{products}
\end{figure}

\noindent
The product of two diagrams is calculated by  drawing the diagrams 
next to each other and joining the right legs of the 
left diagram with the left legs of the right diagram which are 
situated at the same level on the vertical. 
Figure \ref{products} illustrates an example.


\noindent
More formally if $d_i=(\mathcal{V}_i, f_{l,i}, f_{r,i})$ for $i=1,2$ 
with the notations from Definition \ref{def.semigroup}, then 
$d_1\cdot d_2=(\mathcal{V}, f_{l}, f_{r})$ with
\begin{equation}
\mathcal{V}=\mathcal{V}_1\cup\mathcal{V}_2\cup 
\{(f_{r,1}^{-1}(i),f_{l,2}^{-1}(i)):i\leq\mathrm{min}(|R_1|,|L_2|)\},
\end{equation} 
$f_{l}$ is defined on the disjoint union
$L_1+(L_2\setminus f_{l,2}^{-1}(\{1,\dots,\mathrm{min}(|R_1|,|L_2|)\} )$ by
\begin{displaymath}
\left\{ \begin{array}{ll}
 f_l(a)=f_{l,1}(a) & \textrm{for}~ a\in L_{1}\\
 f_l(b)=f_{l,2}(b)+|L_1| & \textrm{for}~ b\in 
L_2\setminus f_{l,2}^{-1}(\{1,\dots,\mathrm{min}(|R_1|,|L_2|)\} 
\end{array}\right.
\end{displaymath} 
and similarly for $f_{r}$. The product does not depend on the chosen 
representatives for $d_i$ in their equivlence class and is associative. 
The diagrams with no legs are the pair partitions, thus 
$\mathcal{P}_2(\infty)\subset\mathcal{B}\mathcal{P}_2(\infty)$. 

\begin{figure}[h]
\begin{center}
\includegraphics[width=10cm]{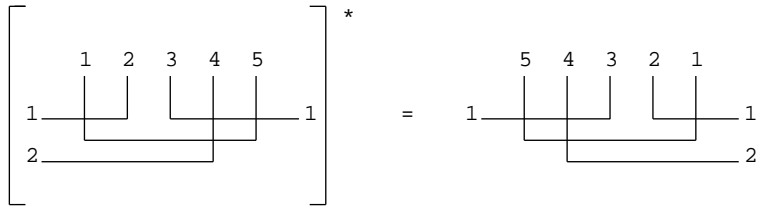}
\end{center}
\caption{The adjoint}
\label{adjoint}
\end{figure}

\noindent
The involution is given by mirror reflection (see Figure \ref{adjoint}). If 
$d=(\mathcal{V},f_l, f_r)$ then $d^*=(\mathcal{V}^*,f_r,f_l)$ 
with the underlying set $X^*$ obtained by reversing the order on $X$ and 
\begin{equation}
\mathcal{V}^* :=\{(b,a):(a,b)\in\mathcal{V}\} 
\end{equation}
is the adjoint of $\mathcal{V}$. It is easy to check that
\begin{displaymath}
(d_1\cdot d_2)^*=d_2^*\cdot d_1^*.
\end{displaymath}   


Let $\mathbf{t}$ be a linear functional on pair partitions. We extend it to 
a function $\hat{\mathbf{t}}$ on $\mathcal{B}\mathcal{P}_2(\infty)$ 
defined as
\begin{equation}\label{def.t.hat}
\hat{\mathbf{t}}(d)=\left\{\begin{array}{ll}
\mathbf{t}(d) & \textrm{if $d\in\mathcal{P}_2(\infty) $} \\
0 & \textrm{otherwise}.
\end{array}\right.
\end{equation}

\begin{theorem}\label{th.semigroup.rep}
The function $\mathbf{t}$ on pair partitions is positive definite 
if and only if $\hat{\mathbf{t}}$ is postive on the 
$^*{-}$semigroup $\mathcal{B}\mathcal{P}_2(\infty)$.
\end{theorem}

\noindent
\textit{Proof.}
The main ideas are already present in the proof of 
Proposition \ref{th.GNSrep}. A GNS-type of construction associates to the pair
$(\mathcal{B}\mathcal{P}_2(\infty), \hat{\mathbf{t}})$
a cyclic representation $\chi_{\mathbf{t}}$ of 
$\mathcal{B}\mathcal{P}_2(\infty)$ on a Hilbert space $V$
with cyclic vector $\xi\in V$. We have 
$\br\xi,\chi_{\mathbf{t}}(d)\xi\ke=\hat{\mathbf{t}}(d)$.
We denote by $\mathcal{B}\mathcal{P}_2^{(n,0)}$ the set of diagrams 
with $n$ left legs and no right legs. Then using 
\begin{equation}
\br\chi_{\mathbf{t}}(d_1)\xi,\chi_{\mathbf{t}}(d_2)\xi\ke_V=
\hat{\mathbf{t}}(d_1^*\cdot d_2)
\end{equation}
we obtain:

\noindent
1. the representation space $V$ is of the form
\begin{equation}
V=\Dsum_{n=0}^\infty V_n \qquad\mathrm{where}\qquad
V_n=\overline{\mathrm{lin}
\{\chi_{\mathbf{t}}(d)\xi:d\in\mathcal{B}\mathcal{P}_2^{(n,0)}\} }
\end{equation}

\begin{figure}[h]
\psfrag{a}{$\tau_{1,2}$}
\begin{center}
\includegraphics[width=10cm]{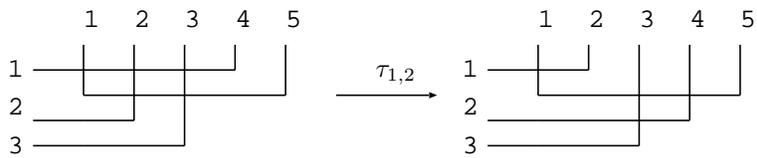}
\end{center}
\caption{The action of a transposition}
\label{transposition}
\end{figure}

\noindent
2. on $\mathcal{B}\mathcal{P}_2^{(n,0)}$ there is an obvious action of 
$\symm (n)$ by permutations of the positions of the left ends of the legs. 
Figure \ref{transposition} shows the action of the transposition $\tau_{1,2}$.
 
 
\noindent
This induces a unitary representation of $\symm (n)$ on $V_n$ as 
\begin{equation}
\tau(d_1)^*\cdot\tau(d_2)=d_1^*\cdot d_2  
\end{equation}
for all $d_1,d_2\in\mathcal{B}\mathcal{P}_2^{(n,0)}$ and $\tau\in\symm (n)$.

\noindent
3. let $d_0\in\mathcal{B}\mathcal{P}_2^{(1,0)}$
 be the ``left hook'' (the diagram with no pairs). 
Then $j:=\chi_{\mathbf{t}}(d_0)$ is an operator on 
$V$ whose restriction $j_n$ to $V_n$ maps it into $V_{n+1}$ 
and satisfies the intertwining condition 
(\ref{intertwining}) with respect ot the 
representations of the symmetric groups on $V_n$ and $V_{n+1}$. 

\noindent
Using the data $(V_n,j_n)$ we construct the triple 
$(\F_V(\h),\mathcal{C}_{V,j}(\h), \Omega_V )$. According to Proposition 
\ref{prop.combinatorics} there exists a positive definite function on
 pair partitions $\mathbf{t}'$ such that $\rho_{V,j}=\rho_{\mathbf{t}'}$. 
We have to prove 
that $\mathbf{t}$, which is the restriction of $\hat{\mathbf{t}}$ to 
$\mathcal{P}_2(\infty)$ coincides with $\mathbf{t}'$. 

\begin{figure}[htbp]
\begin{center}
\includegraphics[width=10cm]{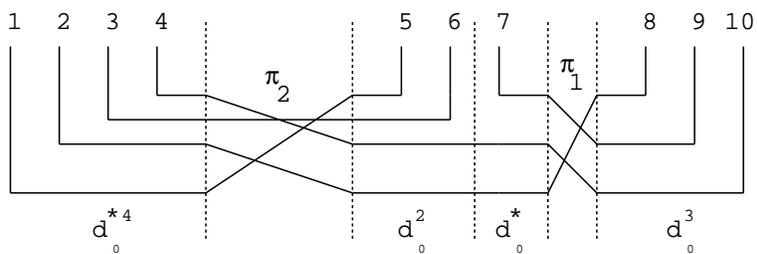}
\end{center}
\caption{The standard form of a pair partition} 
\label{standardform}
\end{figure}

\noindent
Any pair partition $\mathcal{V}$ can be written in a ``standard form''
(see Figure \ref{standardform}): 
\begin{equation}\label{eq.standard.form}
\mathcal{V}= (d_0^*)^{p_m}\cdot\pi_{m-1}( \dots\pi_2( d_0^{k_2} 
\cdot(d_0^*)^{p_1}\cdot\pi_1(d_0^{k_1})))
\end{equation}
where the permutations $\pi_i$ are uniquely defined
by the requirement that any two
lines connecting two pairs in the associated graphic 
intersect minimally and at the rightmost possible position. 


\noindent
Let $\prod_{k=1}^{2n}a_{V,j}^{\sharp_k}(e_{i_k})$ be a monomial containing $n$ 
creation operators and $n$ annihilation operators such that by pairing 
creators with annihilators of the same colour on their right side, 
we generate a pair partition $\mathcal{V}$. The 
definitions (\ref{defcr}), (\ref{defann}) of the 
creation and annihilation operators give their expressions 
in terms of the operator $j, j^*$ and the unitary representations of the 
permutation groups on the spaces $V_n$. By using the intertwining property 
(\ref{intertwining}) we can pass all permutations to the left of the 
$j$-terms and obtain:
\begin{eqnarray}\nonumber
\mathbf{t}'(\mathcal{V}) & = &
\br\Omega_V,\prod_{k=1}^{2n}a_{V,j}^{\sharp_k}(e_{i_k})\Omega_V\ke \\
&= &\br\xi,(j^*)^{p_m}\cdot U(\pi_{m-1}) \dots U(\pi_2)\cdot j^{k_2} 
\cdot (j^*)^{p_1}\cdot U(\pi_1)\cdot j^{k_1}\xi\ke_V \nonumber\\
&= & \br\xi,\chi_{\mathbf{t}}(\mathcal{V}) \xi\ke_V   =
\hat{\mathbf{t}}(\mathcal{V}) \nonumber
\end{eqnarray}
Conversely, starting from a positive definite function $\mathbf{t}$ 
we construct the representation 
$(V, \chi_{\mathbf{t}}(\mathcal{B}\mathcal{P}_2(\infty)),\xi)$ 
through applying Theorem \ref{th.GNSrep} and thus $\hat{\mathbf{t}}$ 
is positive on $\mathcal{B}\mathcal{P}_2(\infty)$.
 
\qed

\section{Generalised Wick products}
\label{gen.wick.products}

 As argued in the introduction, the representations of the 
``field algebras'' $\A(\K)$ with respect to Gaussian
states $\tilde{\rho}_{\mathbf{t}}$ give rise to (noncommutative) processes 
called generalised Brownian motions \cite{Boz.Sp.1} 
for $\K$ (infinite dimensional) real Hilbert space. In all known examples 
such representations appear as restrictions to the subalgebra $\A(\K)$   
of Fock representations
of the algebra of creation and annihilation operators
$\mathcal{C}(\K_\C)$  with respect to the state $\rho_{\mathbf{t}}$. 
We will prove that this is always the case, thus answering a question 
put in \cite{Boz.Sp.1}.

\noindent
Let 
\begin{equation}
\mathbf{t}: \mathcal{P}_2(\infty)\ra \C
\end{equation}
be such that $\tilde{\rho}_{\mathbf{t}}$ is a Gaussian state on $\A(\K)$
 for $\K$ infinite dimensional Hilbert space. Let 
$(\tilde{\mathcal{F}}_{\mathbf{t}}(\K), 
  \tilde{\pi}_{\mathbf{t}}(\A(\K)), \tilde{\Omega}_{\mathbf{t}})$ 
be the GNS-triple associated to $(\mathcal{A}(\K),\tilde{\rho}_{\mathbf{t}})$. 
The $^*$-algebra $\tilde{\pi}_{\mathbf{t}}(\A(\K))$ 
is generated by the  symmetric operators 
$\omega_{\mathbf{t}} (f):=\tilde{\pi}_{\mathbf{t}}(\omega (f))$ 
for all $f\in \K$ with commun domain 
$D:=\tilde{\pi}_{\mathbf{t}}(\mathcal{A}(\K))\tilde{\Omega}_{\mathbf{t}}$. 
The selfadjointness of the field operators will be addressed in section 
\ref{sec.secondquantisation}. For the moment, all operators discussed are 
defined on $D$. \\

In analogy to (\ref{def.secquant.Hilbert}) for any orthogonal operator 
$O\in\mathcal{O}(\K)$ there exists a unitary
\begin{equation}
\tilde{\F}_{\mathbf{t}}(O):
\prod_{k=1}^n \om_{\mathbf{t}}(f_k)\tilde{\Omega}_{\mathbf{t}}  \ra 
\prod_{k=1}^n \om_{\mathbf{t}}(Of_k)\tilde{\Omega}_{\mathbf{t}}
\end{equation}
and 
$\tilde{\F}_{\mathbf{t}}(O_1)\tilde{\F}_{\mathbf{t}}(O_2)=
\tilde{\F}_{\mathbf{t}}(O_1\cdot O_2)$ for 
$O_1,O_2\in\mathcal{O}(\K)$. 
This induces an action on the 
$^*$-algebra $\tilde{\pi}_{\mathbf{t}}(\A(\K))$:  
\begin{equation}
\tilde{\Gamma}_{\mathbf{t}}(O):
X\mapsto \tilde{\F}_{\mathbf{t}}(O)X \tilde{\F}_{\mathbf{t}}(O^*).
\end{equation}

\noindent
Certain operators play a similar role to 
that of the Wick products in quantum field theory \cite{Simon,Str.Wigh.} 
or for the $q$-deformed Brownian motion \cite{Boz.Ku.Spe., Boz.Sp.2}. 
 
\begin{definition}\label{def.Wick.prod} {\rm 
Let $\{P, F\}$ be a partition of the ordered set $\{1,\dots ,2p+n\}$ 
with $|P|=2p$ and $|F|=n$. 
Let $\mathcal{V}=\{(l_1,r_1),\dots ,(l_{p},r_{p})\} \in\mathcal{P}_2(P)$ 
and $\mathbf{f}:F\ra\K$. \\
For every 
$\mathcal{V}'=\{(l'_1,r'_1),\dots ,(l'_{p'},r'_{p'})\}\in\mathcal{P}_2(P')$ 
with $P'\subset F$ we define
\begin{equation}
\eta_\mathbf{f}(\mathcal{V}'):=
\prod_{i=1}^{p'}\br \mathbf{f}(l'_{i}),\mathbf{f}(r'_{i})\ke.
\end{equation}  
The \textit{generalised Wick product} associated to $(\mathcal{V},\mathbf{f})$ 
is the operator $\Psi(\mathcal{V},\mathbf{f})$ determined recursively by 
\begin{eqnarray}
\Psi(\mathcal{V},\mathbf{f}) & + & 
\sum_{\emptyset\neq P'\subset F}~~\sum_{\mathcal{V}'\in \mathcal{P}_2(P')}
\eta_\mathbf{f}(\mathcal{V}')\cdot 
\Psi(\mathcal{V}\cup \mathcal{V}',\mathbf{f}\upharpoonright_{F\setminus P'} )= 
M(\mathcal{V},\mathbf{f})
  \nonumber\\
M(\mathcal{V},\mathbf{f}) & := &  
\mathop{\mathrm{w{-}lim}}_{n\ra \infty}\prod_{k=1}^{2p+n}
\om_{\mathbf{t}}(f_{k,n})
\label{eq.Wick.prod}
\end{eqnarray}
where $f_{k,n}:=\mathbf{f}(k)$ for $k\in F$ and $f_{l_i,n}=
f_{r_i,n}=e_{np+i}$ for $i=1,\dots, p$ with $(e_l)_{l\in\mathbb{N}}$ 
a set of normalised vectors, orthogonal to each other. }
\end{definition}

\noindent
\textbf{Remarks.} 
1) The right side of the last equation needs some clarifications.
The operator $M(\mathcal{V},\mathbf{f})$ 
is defined on $D$ by its matrix elements. 
If 
$\psi_i=
\prod_{a=1}^{m_i}\om_{\mathbf{t}}(g_a^{(i)})\tilde{\Omega}_{\mathbf{t}}$ 
for $i=1,2$ are vectors in $D$ then from the definition of the Gaussian state 
follows immediately that
\begin{equation}
\br\psi_1,M(\mathcal{V},\mathbf{f})\psi_2\ke=
\lim_{n\ra\infty}
\br\psi_1,\prod_{k=1}^{2p+n}\om_{\mathbf{t}}(f_{k,n})\psi_2\ke
\end{equation}

\noindent 
exists and does not depend on the choice of the vectors 
$(e_i)_{i\in\mathbb{N}}$ (as long as they are normal and 
orthogonal to each other) but depends only on their 
positions in the monomial which are determined by the pair partition 
$\mathcal{V}$. In the limit only those pair partitions which contain 
the pairs $(l_i,r_i)\in\mathcal{V}$ give a nonzero contribution.
Thus $M(\mathcal{V},\mathbf{f})$ is well defined.

\noindent
2) If the vectors $(\mathbf{f}(k))_{k=1}^n$ are orthogonal on each other 
then $\eta_\mathbf{f}(\mathcal{V}')=0$, 
thus $\Psi(\mathcal{V},\mathbf{f})=M(\mathcal{V},\mathbf{f})$.

\noindent
3) The dense domain $D$ is spanned by the vectors of the form 
$\Psi(\mathcal{V},\mathbf{f})\tilde{\Omega}_{\mathbf{t}}$. 
Indeed let 
$\psi=
\prod_{k=1}^{n}\om_{\mathbf{t}}(\mathbf{f}(k))\tilde{\Omega}_{\mathbf{t}}$; 
then
\begin{equation}
\psi=\Psi(\emptyset,\mathbf{f})\tilde{\Omega}_{\mathbf{t}}+ 
\sum_{\emptyset\neq P'\subset F}~\sum_{d'\in \mathcal{P}_2(P')}
\eta_f(\mathcal{V}')\cdot 
\Psi(\mathcal{V}',\mathbf{f}\upharpoonright_{F\setminus P'} )
\tilde{\Omega}_{\mathbf{t}}
\end{equation}
with $F=\{1,\dots ,n\}$.

\noindent
4) The choice for $\{1,\dots ,2p+n\}$ as the underlying ordered set is not 
essential. It is useful to think of $\Psi(\mathcal{V},\mathbf{f})$ 
in terms of an arbitrary underlying finite ordered set $X$, 
where $\mathcal{V}\in\mathcal{P}_2(A)$, $A\subset X$, 
$\mathbf{f}: X \setminus A\ra \K$. 
For example we can consider the set 
$X=\{0\}$ and $\mathbf{f}(0)=h$, then 
$\Psi(\emptyset, \mathbf{f})=\om_{\mathbf{t}}(h)$.

\noindent
The relation between $M(\mathcal{V},\mathbf{f})$ and 
$\Psi(\mathcal{V},\mathbf{f})$ 
is similar to the one between moments and cumulants.

\begin{lemma}\label{lemma.moments.cumulants}
Let $\Psi(\mathcal{V},\mathbf{f}),$ $M(\mathcal{V},\mathbf{f})$ 
be as in Definition 
\ref{def.Wick.prod}. 
The equations (\ref{eq.Wick.prod}) can be inverted into:
\begin{equation}
\Psi(\mathcal{V},\mathbf{f})=M(\mathcal{V},\mathbf{f})+
\sum_{\emptyset\neq P'\subset F}~\sum_{\mathcal{V}'\in\mathcal{P}_2(P')}
(-1)^{\frac{|P'|}{2}} \eta_\mathbf{f}(\mathcal{V}')\cdot 
M(\mathcal{V}\cup \mathcal{V}',\mathbf{f}\upharpoonright_{F\setminus P'}).
\end{equation}
\end{lemma}

\noindent
\textit{Proof.} Apply M\"obius inversion formula.

\qed 

\noindent
Let $X$ be an ordered set. Let $\{P,F\}$ be a partition of 
$X$ into disjoint sets and consider a pair 
$(\mathcal{V}\in\mathcal{P}_2(P),\mathbf{f}:F\ra \K)$. 
Then for $X^*$ as underlying set we define the pair 
$(\mathcal{V}^*, \mathbf{f}^*)$ 
where $\mathcal{V}^*\in\mathcal{P}_2(X^*)$ contains the same pairs as 
$\mathcal{V}$ but with the reversed order and $\mathbf{f}^*=\mathbf{f}$. 

\begin{lemma}\label{lemma.adjoint}
With the above notations the following relation holds:
\begin{equation} 
\Psi(\mathcal{V},\mathbf{f})^*=\Psi(\mathcal{V}^*,\mathbf{f}^*).
\end{equation}
\end{lemma}   

\noindent
\textit{Proof.} Apply Lemma \ref{lemma.moments.cumulants} and use 
$M(\mathcal{V},\mathbf{f} )^*=M(\mathcal{V}^*,\mathbf{f}^*)$ 
which follows directly from Definition \ref{def.Wick.prod}.

\qed

\noindent
For two ordered sets $X$ and $Y$ we define their concatenation 
$X+Y$ as the disjoint union with the original order on $X$ and $Y$
 and with $x<y$ for any $x\in X, y\in Y$. If $\mathbf{f}_X:X\ra\K$ and 
$\mathbf{f}_Y:Y\ra \K$ then we denote by 
$\mathbf{f}_X\oplus \mathbf{f}_Y$ the function on 
$X+Y$ which restricts to $\mathbf{f}_X$ and $\mathbf{f}_Y$ on $X$ respectively $Y$. 
Finally if $|X|=|Y|=m$ we identify the subset of $\mathcal{P}_2(X+Y)$:
\begin{equation}
\mathcal{P}_2(X,Y):=
\{\{(x_1,y_1),\dots ,(x_m, y_m)\}~:~x_i\in X, y_i\in Y, i=1,\dots ,m\}
\end{equation}  
\begin{lemma}
\label{lemma.inn.prod.W} 
Let $(P_i,F_i)$ be a disjoint partition of $X_i$ and 
$\mathcal{V}_i\in \mathcal{P}_2(P_i)$, 

\noindent
$\mathbf{f}_i:F_i\ra \K$ for $i=1,2$.  Then
\begin{equation}
\br \Psi(\mathcal{V}_1, \mathbf{f}_1)\tilde{\Omega}_{\mathbf{t}},~
    \Psi(\mathcal{V}_2, \mathbf{f}_2)\tilde{\Omega}_{\mathbf{t}}\ke =
\delta_{|F_1|, |F_2|} 
\sum_{\mathcal{V}\in \mathcal{P}_2(F_1^*,F_2)}
\eta_{\mathbf{f}_1^*\oplus \mathbf{f}_2}(\mathcal{V})\cdot 
\mathbf{t}(\mathcal{V}_1^*\cup \mathcal{V}_2\cup \mathcal{V})
\end{equation}
with the convention 
$\eta_{\mathbf{f}_1^*\oplus \mathbf{f}_2}(\mathcal{V})=1$ for 
$F_1=F_2=\emptyset$.
\end{lemma} 

\noindent
\textit{Proof.} From Definitions \ref{def.Gaussian.state}, 
\ref{def.Wick.prod} it follows that
\begin{equation}\label{eq.inn.prod.M}
\br M(\mathcal{V}_1, \mathbf{f}_1)\tilde{\Omega}_{\mathbf{t}},~
    M(\mathcal{V}_2, \mathbf{f} _2)\tilde{\Omega}_{\mathbf{t}}\ke =
\sum_{\mathcal{V}\in \mathcal{P}_2(F_1^*+F_2)}
\eta_{\mathbf{f}_1^*\oplus \mathbf{f}_2}(\mathcal{V})\cdot 
\mathbf{t}(\mathcal{V}_1^*\cup \mathcal{V}_2\cup \mathcal{V}).
\end{equation}
We apply Lemma \ref{lemma.moments.cumulants} and obtain:
\begin{eqnarray}
&& \br \Psi(\mathcal{V}_1, \mathbf{f}_1)\tilde{\Omega}_{\mathbf{t}},~
       \Psi(\mathcal{V}_2, \mathbf{f}_2)\tilde{\Omega}_{\mathbf{t}}\ke =
\sum_{\mathcal{V}_1',\mathcal{V}_2'}(-1)^{\frac{|P'_1|+|P'_2|}{2}}\cdot 
\eta_{\mathbf{f}_1^*\oplus \mathbf{f}_2}(\mathcal{V}_1^{\prime *}\cup\mathcal{V}_2')
  \cdot \nonumber\\
&&\cdot
\br M(\mathcal{V}_1\cup \mathcal{V}_1', 
\mathbf{f}_1\upharpoonright_{F_1\setminus P'_1})\tilde{\Omega}_{\mathbf{t}},~
M(\mathcal{V}_2\cup \mathcal{V}_2', 
\mathbf{f}_2\upharpoonright_{F_2\setminus P'_2})\tilde{\Omega}_{\mathbf{t}}\ke
            \nonumber
\end{eqnarray}
where the sum runs over all 
$\mathcal{V}_i'\in\mathcal{P}_2(P'_i), P'_i\subset F_i$ for $i=1,2$. 
Substituting in the last expression the result from equation 
(\ref{eq.inn.prod.M}) it becomes:
\begin{equation}
\sum_{\mathcal{V}_1',\mathcal{V}_2'}
\sum_{\mathcal{V}}(-1)^{\frac{|P'_1|+|P'_2|}{2}}\cdot
\eta_{\mathbf{f}_1^*\oplus \mathbf{f}_2}(\mathcal{V}_1^{\prime *}\cup
\mathcal{V}_2'\cup\mathcal{V} )\cdot  
\mathbf{t}((\mathcal{V}_1\cup \mathcal{V}_1')^*\cup 
\mathcal{V}_2\cup\mathcal{V}_2'\cup \mathcal{V})
\end{equation}
with the second sum running over all 
$\mathcal{V}\in \mathcal{P}_2((F_1\setminus P'_1)^* +(F_2\setminus P'_2))$. 
We make the notation 
$\tilde{\mathcal{V}}:=
\mathcal{V} _1^{\prime *}\cup\mathcal{V}_2'\cup\mathcal{V} $ 
and by grouping together all terms containing $\tilde{\mathcal{V}}$ 
the initial expression looks like:
\begin{equation}
\br \Psi(\mathcal{V}_1, \mathbf{f}_1)\tilde{\Omega}_{\mathbf{t}},~
    \Psi(\mathcal{V}_2, \mathbf{f}_2)\tilde{\Omega}_{\mathbf{t}}\ke =
\sum_{\tilde{\mathcal{V}}}
m(\tilde{\mathcal{V}})
\cdot\eta_{\mathbf{f}_1^*\oplus \mathbf{f}_2}(\tilde{\mathcal{V}})
\cdot\mathbf{t}(\mathcal{V}_1^*\cup\tilde{\mathcal{V}}\cup \mathcal{V}_2)
\end{equation}
where the symbol $m(\tilde{\mathcal{V}})$ stands for total 
contribution from the terms of the form $(-1)^{\frac{|P'_1|+|P'_2|}{2}}$. 
We calculate now $m(\tilde{\mathcal{V}})$:
\begin{equation}
m(\tilde{\mathcal{V}})=\sum_{\mathcal{V}_1',\mathcal{V}_2',\mathcal{V}}
(-1)^{|\mathcal{V}_1'|+|\mathcal{V}_2'|},
\end{equation}  
this sum running over all 
$\mathcal{V}\in \mathcal{P}_2((F_1\setminus P'_1)^* +(F_2\setminus P'_2))$, 
$\mathcal{V}_i'\in\mathcal{P}_2(P'_i)$, $P'_i\subset F_i$ 
for $i=1,2$ with the constraint 
$\tilde{\mathcal{V}}=
\mathcal{V}_1^{\prime *}\cup \mathcal{V}_2'\cup\mathcal{V}$. 

\noindent
Suppose that 
$\tilde{\mathcal{V}}\in\mathcal{P}_2(F_1^* ,F_2)$, then 
$\mathcal{V}_1'=\mathcal{V}_2'=\emptyset$ and 
$m(\tilde{\mathcal{V}})=1$. Otherwise $\tilde{\mathcal{V}}$ 
can be written in a unique way as 
\begin{equation}
\tilde{\mathcal{V}}=
\tilde{\mathcal{V}}_1^*\cup\tilde{\mathcal{V}}_c\cup\tilde{\mathcal{V}}_2
\end{equation}
where $\tilde{\mathcal{V}}_i\in\mathcal{P}_2(\tilde{P}_i)$, 
$\emptyset\neq\tilde{P}_i\subset X_i$ 
for $i=1,2$ and 
$\tilde{\mathcal{V}}_c\in\mathcal{P}_2
((X_1\setminus \tilde{P}_1)^*,X_2\setminus \tilde{P}_2)$. 
Then one has the inclusions $\mathcal{V}_i'\subset\tilde{\mathcal{V}}_i$ 
for $i=1,2$ and $\mathcal{V}_c\subset\mathcal{V} $. 
The calculation of $m(\tilde{\mathcal{V}})$ reduces then to 
\begin{equation}
m(\tilde{\mathcal{V}})=\sum_{\mathcal{V}_1'\subset
\tilde{\mathcal{V}}_1,\mathcal{V}_2'\subset\tilde{\mathcal{V}}_2}
(-1)^{|\mathcal{V}_1'|+|\mathcal{V}_2'|}
=(1-1)^{|\tilde{\mathcal{V}}_1|+|\tilde{\mathcal{V}}_2|}=0.
\end{equation} 
In conclusion 
\begin{equation}
\br \Psi(\mathcal{V}_1, \mathbf{f}_1)\tilde{\Omega}_{\mathbf{t}},~
    \Psi(\mathcal{V}_2, \mathbf{f}_2)\tilde{\Omega}_{\mathbf{t}}\ke =
\sum_{\tilde{\mathcal{V}}\in \mathcal{P}_2(F_1^*,F_2)}
\eta_{\mathbf{f}_1^*\oplus \mathbf{f}_2}(\tilde{\mathcal{V}})\cdot 
\mathbf{t}(\mathcal{V}_1^*\cup \mathcal{V}_2\cup \tilde{\mathcal{V}})
\end{equation}     

\qed

A similar result holds for algebras of creation 
and annihilation operators. 
Suppose that $\mathbf{t}$ is a function (not neccesarily positive definite) 
on pair partitions. Let $P,F,\mathcal{V} , \mathbf{f}$ be as in 
Definition \ref{def.Wick.prod} and define in the representation space 
$\mathcal{F}_{\mathbf{t}}(\K_\C)$ the vectors 
\begin{equation}\label{def.psi.vectors}
\psi(\mathcal{V},\mathbf{f})=
\prod_{k=1}^{2p+n}a^{\sharp_k}_{\mathbf{t}}(f_k))\Omega_{\mathbf{t}}
\end{equation}
with $a^{\sharp_k}(f_k)=a^*(\mathbf{f}(k))$ for $k\in F$,  
$a^{\sharp_{l_i}}(f_{l_i})=(a^{\sharp_{r_i}}(f_{r_i}))^*=a(g_i)$ 
 for $i=1,\dots ,p$ and $(g_i)_{i=1,\dots ,p}$ a set of normalised vectors, 
orthogonal to each other and to the vectors $(\mathbf{f}(k))_{k=1}^n$. 

\begin{lemma}\label{lemma.inn.prod}
Let $\mathbf{t}$ be a function on pair partitions. Then
\begin{equation}\label{eq.inn.prod}
\br \psi (\mathcal{V}_1, \mathbf{f}_1),~
    \psi (\mathcal{V}_2, \mathbf{f}_2)
\ke_{\mathcal{F}_{\mathbf{t}}(\K_\C)} =
\sum_{\mathcal{V}\in \mathcal{P}_2(F_1^*,F_2)}
\eta_{\mathbf{f}_1^*\oplus \mathbf{f}_2}(\mathcal{V})\cdot 
\mathbf{t}(\mathcal{V}_1^*\cup \mathcal{V}_2\cup\mathcal{V} )
\end{equation}
\end{lemma}

\noindent 
\textit{Proof.}
The equation follows then directly from 
Definition \ref{def.Fock.state}.

\qed

\noindent
Now we are ready for the main result of this section.
 
\begin{theorem}\label{th.gauss.vs.fock}
Let $\mathbf{t}$ be a function on pair partitions. 
If $\tilde{\rho}_{\mathbf{t}}$ 
is a Gaussian state on $\mathcal{A}(\K)$ for any real Hilbert space $\K$ then 
$\rho_{\mathbf{t}}$ is a Fock state on $\mathcal{C}(\K_{\C})$.
\end{theorem}

\noindent 
\textit{Proof.}
Suppose that $\rho_{\mathbf{t}}$ is not a Fock state. Then in the 
representation space 
$\mathcal{F}_{\mathbf{t}}(\K_\C)$ there exists a vector of the form
\begin{equation}
\psi=\sum_{a=1}^m c_a\cdot\psi(\mathcal{V}_a, \mathbf{f}_a)   
\end{equation}
with all $\mathbf{f}_a$ taking values in the real subspace 
$\K$ of $\K_\C$ and $c_a\in \C$, such that $\br\psi,\psi\ke<0$. 
But from lemmas \ref{lemma.inn.prod.W} and 
\ref{lemma.inn.prod} it results that \\
$\|\sum_{a=1}^m c_a\cdot \Psi(\mathcal{V}_a, \mathbf{f}_a)
\tilde{\Omega}_{\mathbf{t}}\|^2<0$ which is
 a contradiction. Thus  $\rho_{\mathbf{t}}$ is a positive functional and 
$\mathbf{t}$ is a positive definite function on pair partitions.

\qed

\noindent
From Lemmas \ref{lemma.moments.cumulants} and \ref{lemma.adjoint}  
we conclude that the generalised Wick products  
$\Psi(\mathcal{V},\mathbf{f})$ acting on $\mathcal{F}_{\mathbf{t}}(\K)$ 
form a $^*$-algebra of operators which contains 
$\pi_{\mathbf{t}}(\mathcal{A}(\K))$ and will be denoted 
by $\tilde{\Delta}_{\mathbf{t}}(\K)$.  Let us first note that Theorem 
\ref{th.gauss.vs.fock} implies that the representations of 
$\tilde{\Delta}_{\mathbf{t}}(\K)$ 
on $\mathcal{F}_{\mathbf{t}}(\K_\C)$ and 
$\tilde{\mathcal{F}}_{\mathbf{t}}(\K)$ are unitarily equivalent, thus:
\begin{corollary}\label{cor.wick.cyclic}
The vacuum vector $\Omega_{\mathbf{t}}$ is cyclic for the $^*$-algebra 
$\tilde{\Delta}_{\mathbf{t}}(\K)$ for any real Hilbert space $\K$.
\end{corollary}

\section{Second Quantisation}
\label{sec.secondquantisation}

This section is dedicated to the description of functorial properties 
of the generalised Brownian motion which go by the name of 
second quantisation and appear at 
two different levels depending on the categories with which we work.

\noindent
Let $\h, ~\h'$ be  Hilbert spaces and $T$ a 
contraction from $\h$ to $\h'$. Define the second quantisation 
of $T$ at the Hilbert space level by
\begin{eqnarray}
\mathcal{F}_{\mathbf{t}}(T): \mathcal{F}_{\mathbf{t}}(\h) & \ra & 
\mathcal{F}_{\mathbf{t}}(\h')\nonumber\\
v\tens_s h_0\tens\dots\tens h_{n-1} & \mapsto & 
v\tens_s Th_0\tens\dots\tens Th_{n-1}
\end{eqnarray}
for all $v\in V_n$, $h_i\in \h$ when $n \geq 1$, 
and equal to the identity on $V_0$. Clearly $\mathcal{F}_{\mathbf{t}}(T)$ 
is a contraction, satisfies the equation
$\mathcal{F}_{\mathbf{t}}(T_1)\cdot\mathcal{F}_{\mathbf{t}}(T_2)=\mathcal{F}_{\mathbf{t}}(T_1\cdot T_2)$ 
and for $T$ unitary it coincides with the operator defined
in the equations (\ref{defsecquant}) and (\ref{defF_V}). 

\begin{definition}{\rm
We call $\mathcal{F}_{\mathbf{t}}$ the functor of 
\textit{second quantisation at the Hilbert space level} 
. }
\end{definition}

\begin{lemma}
Let $\psi(\mathcal{V},\mathbf{f})$ as defined in equation 
(\ref{def.psi.vectors}). Then
\begin{equation}
\mathcal{F}_{\mathbf{t}}(T)\psi(\mathcal{V},\mathbf{f})=
                           \psi(\mathcal{V},T\circ \mathbf{f}).
\end{equation}
\end{lemma}

\noindent
\textit{Proof.} We use the representation $\chi_{\mathbf{t}}$ of the 
$^*$-semigroup of broken pair partitions 
$\mathcal{B}\mathcal{P}_2(\infty)$ with respect to the state 
$\hat{\mathbf{t}}$ (see equation \ref{def.t.hat}). 
Let $\{F,P\}$ be a partition of 
$\{1,\dots ,2p+n\}$ and $\mathcal{V}\in\mathcal{P}_2(P)$, 
$\mathbf{f}:F\ra \h$. Then using (\ref{def.psi.vectors}) and the equations 
(\ref{defcr}, \ref{defann}) we obtain 
\begin{equation}
\psi(\mathcal{V},\mathbf{f})=
\chi_{\mathbf{t}}(\tilde{\mathcal{V}})\xi\tens_s \Tens_{k\in F}\mathbf{f}(k)
\end{equation}
for $\tilde{\mathcal{V}}\in\mathcal{B}\mathcal{P}_2^{n,0}$ 
the diagram with the set of
pairs $\mathcal{V}$ and $n$ legs to the left which do not 
intersect each other.
 
\qed

\noindent
There is however a more interesting notion of second quantisation.  

\begin{definition}{\rm i) 
The category of \textit{non-commutative probability spaces} 
 has as objects pairs $(\mathcal{A},\rho_{\mathcal{A}})$ 
of von Neumann algebras and normal states
and as morphisms between two objects $(\mathcal{A},\rho_{\mathcal{A}})$ and 
 $(\mathcal{B},\rho_{\mathcal{B}})$ all completely positive maps 
$T:\mathcal{A}\rightarrow\mathcal{B}$ such that 
$T(\mathbf{1}_\mathcal{A})=\mathbf{1}_\mathcal{A}$ and 
$\rho_{\mathcal{B}}(Tx)=\rho_{\mathcal{A}}(x)$
for all $x\in\mathcal{A}$.

\noindent 
ii) A functor $\Gamma$ from the category of (real) Hilbert spaces 
with contractions to the category 
of non-commutative probability spaces 
is called \textit{functor of white noise} if $\Gamma(\{0\})=\C$ 
where $\{0\}$ stands for the zero dimensional Hilbert space and satisfies
the continuity requirement
\begin{equation}\label{continuity}
\mathop{\mathrm{w{-}lim}}_{n\ra \infty} 
\Gamma (T_n)(X)= \Gamma (T)(X).
\end{equation}
for any sequence of contractions $T_n:\K\to\K'$  converging weakly to $T$. }
\end{definition}

\noindent
This definition is similar the one in \cite{Kumm.} apart from the 
continuity condition. For completeness we include the following 
standard result.

\begin{proposition}\label{prop.kumm}
If $\Gamma$ is a functor of white noise
then $\Gamma(T)$ is an injective $^*$-homomorphism (automorphism)  
if $T$ is an (invertible) isometry , 
and $\Gamma(P)$ is a conditional expectation if $P$ is an 
orthogonal projection.
\end{proposition}

\noindent
\textit{Proof.}
For separating vacuum the proof has been given in \cite{Kumm.2}. 
Here we do not assume this property.

\noindent
1. Let $O:\K\to \K'$ be an orthogonal operator and 
$X\in\Gamma(\K)$. As $\Gamma(O^*)$ and 
$\Gamma(O)$ are completely positive we have the inequalities 
\begin{equation}\label{eq.ineq1}
\Gamma(O^*)(\Gamma(O)(X^*)\cdot
\Gamma(O)(X))\geq 
\Gamma(O^*O)(X^*)\cdot
\Gamma(O^*O)(X)=X^*X
\end{equation}  
and
\begin{equation}
\Gamma(O)(X^*)\Gamma(O)(X)\leq
\Gamma(O)(X^*X)
\end{equation} 
which by applying the positive operator $\Gamma(O^*)$ becomes
\begin{equation}\label{eq.ineq2}
\Gamma(O^*) 
(\Gamma(O)(X^*)\cdot\Gamma(O)(X))\leq
 \Gamma(O^*O)(X^*X)=X^*X
\end{equation}
From (\ref{eq.ineq1}, \ref{eq.ineq2}) we get 
$\Gamma(O)(X^*)\cdot\Gamma(O)(X)=
\Gamma(O)(X^*X)$ 
and by repeating the argument for $X+Y$ and $X+iY$ we obtain that 
$\Gamma(O)$ is a $^*$-isomorphism. 

\noindent
2. Let $\K$ be a real Hilbert space and $I:\K\to \K\oplus\ell^2(\Z)$ 
the natural isometry. Let $S$ be the shift operator on $\ell^2(\Z)$.
The operator $O:=\mathbf{1}\oplus S$ is orthogonal and 
$\mathop{\mathrm{w{-}lim}}_{n\ra \infty}O^n=P$ where $P$ is the projection 
on $\K$. By the continuity assumption \ref{continuity} we have then
\begin{equation}
\mathop{\mathrm{w{-}lim}}_{n\ra \infty}\Gamma(O^n)(X)=
\Gamma(P)(X)
\end{equation} 
for all $X\in\Gamma(\K\oplus\ell^2(\Z))$. Let 
$\Gamma(\K\oplus\ell^2(\Z))^{\Z}$ be the subalgebra of\\
$\Gamma(\K\oplus\ell^2(\Z))$ of operators invariant under the 
action of the group of automorphisms $(\Gamma(O^n))_{n\in\Z}$. 
Then from $O^nP=P$ we get $\Gamma(P)(X)\in\Gamma(\K\oplus\ell^2(\Z))^{\Z}$. For arbitrary 
$Y_1,Y_2\in\Gamma(\K\oplus\ell^2(\Z))^{\Z}$ the following holds 
\begin{equation}
\Gamma(P)(Y_1XY_2)=Y_1\Gamma(P)(X)Y_2
\end{equation} 
which means that $\Gamma(P)$ is a conditional expectation from
$\Gamma(\K\oplus\ell^2(\Z))$ onto 
$\Gamma(\K\oplus\ell^2(\Z))^\Z$. We show now that 
$\Gamma(I)$ is an injective $^*$-homomorphism. 
By a similar argument to that used in (\ref{eq.ineq1}, \ref{eq.ineq2}) we have:
\begin{equation}\label{eq.I*}
\Gamma(I^*)
(\Gamma(I)(X)\cdot\Gamma(I)(Y))=XY
\end{equation} 
for all $X,Y\in\Gamma(\K)$. Now for any 
$Z\in\Gamma(\K)$
\begin{equation}
\Gamma(I)(Z)=
\Gamma(PI)(Z)\in\Gamma(\K\oplus\ell^2(\Z))^\Z
\end{equation} 
which together with (\ref{eq.I*}) implies 
\begin{equation}
\Gamma(I)(XY)=
\Gamma(II^*)
(\Gamma(I)(X)\cdot\Gamma(I)(Y))=
\Gamma(I)(X)\cdot\Gamma(I)(Y)
\end{equation} 
and thus $\Gamma(I)$ is an injective $^*$-homomorphism. 

\noindent
3. Let $I:\K\to \K'$ be an isometry. We consider the natural isometries
$I_\K':\K'\to\K'\oplus \ell^2(\Z)$ and $I_\K=I_{\K'} I$. From the previous 
argument we know that $\Gamma(I_{\K'}),\Gamma(I_\K)$
are injective $^*$-homomorphisms. Let $X,Y\in\Gamma(\K)$. Then
\begin{eqnarray}
\Gamma(I_{\K'}) (\Gamma(I)(XY))
&=&\Gamma(I_\K)(XY) 
=\Gamma(I_\K)(X)\cdot\Gamma(I_\K)(Y)\nonumber\\
&= &
\Gamma(I_{\K'})(\Gamma(I)(X)\cdot
                           \Gamma(I)(Y)).
\end{eqnarray} 
As $\Gamma(I_{\K'})$ is injective $^*$-homomorphism we obtain 
\begin{equation}
\Gamma(I)(XY)=
\Gamma(I)(X)\Gamma(I)(Y). 
\end{equation}
Thus $\Gamma(I)$ is a $^*$-homomorphism. 
The injectivity follows from $I^*I=\mathbf{1_\K}$.

\noindent
4. Using the previous step of the proof we see that $\Gamma(P)$  
is a norm one projection from $\Gamma(\K')$ onto its von Neumann 
subalgebra $\Gamma(I)(\Gamma(\K))$. Thus 
$\Gamma(P)$ is a conditional expectation \cite{Ta.}.

\qed

\begin{corollary}\label{cor.isomorphism} If $\Gamma$ is a functor of second quantisation
then for any real Hilbert space $\h$ and any infinite dimensional 
real Hilbert space $\K$ the algebras
$\Gamma(\h\oplus\K)^{\mathcal{O}(\K)}$ and $\Gamma(\h)$ are isomorphic
, in particular $\Gamma(\K)^{\mathcal{O}(\K)}=\C \mathbf{1}$.
\end{corollary}

\noindent
\textit{Proof.}
We can choose $\K=\ell^2(\Z)$. Let $S$ be the right shift on 
$\ell^2(\Z)$ and $O=\mathbf{1}\oplus S$ orthogonal operator on $\h\oplus\K$.
The argument given in the previous proposition implies that 
$\Gamma(\h\oplus\K)^{\mathcal{O}(\K)}$ 
is isomorphic with $\Gamma(I)\Gamma(\h)$ where $I$ is 
the natural isometry from  $\h$ to $\h\oplus \K$. Thus 
$\Gamma(\h\oplus\K)^{\mathcal{O}(\K)}\simeq\Gamma(\h)$.

\qed

\noindent
After these general considerations we come back to our construction 
from the previous section: for a fixed positive definite function $\mathbf{t}$ 
we have associated to each Hilbert space $\K$ an algebra 
$\pi_{\mathbf{t}}(\mathcal{A}(\K))$ acting on 
$\mathcal{F}_{\mathbf{t}}(\K_\C)$ and a positive functional 
$\br\Omega_{\mathbf{t}},\cdot \Omega_{\mathbf{t}}\ke$ on the algebra. 
We would like to transform this correspondence into a functor of white noise. 
The natural way to do this is to construct the von Neumann algebra 
generated by the spectral projections of the selfadjoint field operators 
$\omega_{\mathbf{t}}(f)$ for all $f\in\K$. 
However these operators are in general only symmetric and, unless bounded, 
one has to make sure that they are essentially selfadjoint. 
Let us suppose for the moment that this is the case. 
Then we identify two candidates for the image objects under the 
functor of white noise associated to $\mathbf{t}$:\newline

\noindent
1) $\tilde{\Gamma}_{\mathbf{t}}:
\K\mapsto(\tilde{\Gamma}_{\mathbf{t}}(\K), \br\Omega_{\mathbf{t}},\cdot
\Omega_{\mathbf{t}}\ke)$ where $\tilde{\Gamma}_{\mathbf{t}}(\K)$ 
is the von Neumann algebra generated by all the spectral projections 
of the (closed) field operators $\omega_{\mathbf{t}}(f)$ acting on 
$\mathcal{F}_{\mathbf{t}}(\K_\C)$ for all $f\in\K$. \newline

\noindent
2) $\Gamma_{\mathbf{t}}:
\K\mapsto (\Gamma_{\mathbf{t}}(\K), \br\Omega_{\mathbf{t}},\cdot
\Omega_{\mathbf{t}}\ke)$ where $\Gamma_{\mathbf{t}}(\K)$ 
is the von Neumann subalgebra of 
$\tilde{\Gamma}_{\mathbf{t}}(\K\oplus \ell^2(\Z))$ 
consisting of operators which commute with the unitaries 
$\mathcal{F}_{\mathbf{t}}(\mathbf{1}\oplus O)$ for all 
$O\in\mathcal{O}(\ell^2(\Z))$, i.e.
\begin{equation}\label{eq.gamma.k}
\Gamma_{\mathbf{t}}(\K):=
\tilde{\Gamma}_{\mathbf{t}}(\K\oplus \ell^2(\Z))^{\mathcal{O}(\ell^2(\Z))}.
\end{equation}

\noindent
In the cases known so far $-$ the gaussian functor \cite{Simon}, 
the free white noise \cite{Voi.Dy.Ni.} and the $q$-deformed Brownian motion 
\cite{Boz.Ku.Spe.} $-$ the two definitions are equivalent. In fact 
corollary \ref{cor.isomorphism} implies that if 
$\tilde{\Gamma}_{\mathbf{t}}$ is a functor of white noise and the 
isomorphisms $\tilde{\Gamma}_{\mathbf{t}}(O)$ are given by
\begin{equation}
\tilde{\Gamma}_{\mathbf{t}}(O):X\mapsto
\mathcal{F}_{\mathbf{t}}(O) X 
\mathcal{F}_{\mathbf{t}}(O^*)
\end{equation}
for all orthogonal operators $O:\K\ra\K'$ and all
$X\in\tilde{\Gamma}_{\mathbf{t}}(\K)$,
then the algebras $\tilde{\Gamma}_{\mathbf{t}}(\K)$ and 
$\Gamma_{\mathbf{t}}(\K)$ are isomorphic. 
For a general treatement it appears however that 
$\Gamma_{\mathbf{t}}$ is the appropriate definition to start with. 
As above, for any orthogonal operator 
$O:\K\rightarrow\K'$, the natural choice for $\Gamma_{\mathbf{t}}(O)$ is
\begin{equation}\label{eq.sec.quant.orthogonal}  
\Gamma_{\mathbf{t}}(O)(X)=
\mathcal{F}_{\mathbf{t}}(O\oplus\mathbf{1}) X 
\mathcal{F}_{\mathbf{t}}(O\oplus\mathbf{1})^*
\end{equation}
where $X\in\Gamma_{\mathbf{t}}(\K)$. Our task is now to find for 
which functions $\mathbf{t}$ one can construct 
such von Neumann algebras, i.e. the field operators are selfadjoint, 
and moreover the map 
$\K\rightarrow \Gamma_{\mathbf{t}}(\K)$ can be enriched with the morphisms  
\begin{displaymath}
\Gamma_{\mathbf{t}}(T):
\Big(\Gamma_{\mathbf{t}}(\K), \br\Omega_{\mathbf{t}},\cdot \Omega_{\mathbf{t}}\ke\Big)\rightarrow
\Big(\Gamma_{\mathbf{t}}(\K'), \br\Omega_{\mathbf{t}},\cdot \Omega_{\mathbf{t}}\ke\Big)
\end{displaymath}
for all contractions $T:\K\rightarrow\K'$ such that $\Gamma_{\mathbf{t}}$
is a functor of white noise. 

\begin{definition}{\rm
A functor $\Gamma_{\mathbf{t}}$ with the above properties will be called 
\textit{second quantisation at algebraic level} and the completely 
positive map $\Gamma_{\mathbf{t}}(T)$ 
the second quantisation of the contraction $T$. } 
\end{definition} 

The existence of the second quantisation at algebraic level turns 
out to be connected to a property of the functions on pair partitions.

\begin{definition}\label{def.multiplicative}{\rm \cite{Boz.Sp.1} 
A function $\mathbf{t}$ on pair partitions is  called 
\textit{multiplicative} if 
for all $k,l,n\in\N$ with $0 \leq k<l\leq n$ and all 
$\mathcal{V}_1\in\mathcal{P}_2(\{1,\dots ,k,l+1,\dots ,n\})$ and 
$\mathcal{V}_2\in\mathcal{P}_2(\{k+1,\dots ,l\})$ we have
\begin{equation}
\mathbf{t}(\mathcal{V}_1\cup\mathcal{V}_2)=
\mathbf{t}(\mathcal{V}_1)\cdot \mathbf{t}(\mathcal{V}_2).
\end{equation}
}
\end{definition}

\begin{corollary}\label{prop.mutilplicative}
Let $\mathbf{t}$ be a positive definite function on pair 
partitions and suppose that there exists a functor of second quantisation 
$\Gamma_{\mathbf{t}}$. 
Then $\mathbf{t}$ is multiplicative.
\end{corollary} 

\noindent
\textit{Proof.}
Let $\K=\ell^2(\Z)\oplus\ell^2(\Z)$ with the two right shifts 
$S_1,S_2$ acting separately on the two $\ell^2(\Z)$. 
Let $\mathcal{V}_1\cup \mathcal{V}_2$ be a pair partition as in 
Definition \ref{def.multiplicative}. For any pair partition $\mathcal{V}$ 
the operator $\Psi(\mathcal{V},\emptyset)$ on 
$\mathcal{F}_{\mathbf{t}}((\K\oplus\ell^2(\Z))_\C )$ commutes with 
$\mathcal{F}_{\mathbf{t}}(O)$ for all 
$O\in\mathcal{O}(\K\oplus\ell^2(\Z))$. The monomials of field operators 
and the generalised Wick products are affiliated to $\Gamma_{\mathbf{t}}(\K)$. 
By corollary \ref{cor.isomorphism} we have 
\begin{equation}
\Psi(\mathcal{V}, \emptyset)=
\br\Omega_{\mathbf{t}},\Psi(\mathcal{V}, \emptyset)\Omega_{\mathbf{t}}\ke
\mathbf{1}=
\mathbf{t}(\mathcal{V})\mathbf{1}.
\end{equation}
We consider a monomial of fields 
$M(\mathcal{V}_1\cup\mathcal{V}_2)$ 
containing $|\mathcal{V}_1|+|\mathcal{V}_2|$ pairs of different 
colours arranged according to the pair partition 
$\mathcal{V}_1\cup\mathcal{V}_2$ and such that the colours for the pairs in 
$\mathcal{V}_1$ belong to the first $\ell^2(\Z)$ in $\K$, and those 
for the pairs in $\mathcal{V}_2$ belong to the second term. Then  
\begin{eqnarray} 
&& \Psi(\mathcal{V}_1\cup\mathcal{V}_2, \emptyset)=
\mathbf{t}(\mathcal{V}_1\cup \mathcal{V}_2) \mathbf{1}=
 \mathop{\mathrm{w{-}lim}}_{n,m\ra \infty}
 \Gamma_\mathbf{t}(S_1^nS_2^m)(M(\mathcal{V}_1\cup\mathcal{V}_2))\nonumber\\
&&=\mathop{\mathrm{w{-}lim}}_{m\ra \infty}M(\mathcal{V}_1)
\mathbf{t}(\mathcal{V}_2)=
\Psi(\mathcal{V}_1,\emptyset)\cdot \mathbf{t}(\mathcal{V}_2)=
\mathbf{t}(\mathcal{V}_1)\mathbf{t}(\mathcal{V}_2)\cdot\mathbf{1}. 
\end{eqnarray}

\qed

\begin{lemma} Let $\mathbf{t}$ be multiplicative positive definite function. 
Then the operator $j:=\chi_{\mathbf{t}}(d_0)$ defined in 
Theorem \ref{th.semigroup.rep} is an isometry.
\end{lemma}

\noindent
\textit{Proof.} We have
\begin{displaymath}
\br \chi_{\mathbf{t}}(d_1)\xi, j^*j \chi_{\mathbf{t}}(d_1)\xi\ke =
\hat{\mathbf{t}}(d_1^* \cdot p \cdot d_2)=
\hat{\mathbf{t}}(d_1^*d_2)\cdot 1=
\br \chi_{\mathbf{t}}(d_1)\xi, \chi_{\mathbf{t}}(d_1)\xi\ke 
\end{displaymath}
where $p=d_0^* d_0$ is the diagram consisting of one pair and 
$\mathbf{t}(p)=1$ by the normalisation convention in the definition of 
$\mathbf{t}$.  

\qed

\begin{proposition} 
Let $\mathbf{t} $ be multiplicative positive definite function and 
$\psi_k\in\mathcal{F}_{\mathbf{t}}^{(k)}(\K)$ a $k-$particles vector. Then
\begin{equation}
\|\omt (f_1)\dots \omt (f_n)\psi_k \|\leq 
2^{\frac{n}{2}}\sqrt{(k+1)\dots (k+n)} \|\psi_k \| \prod_{i=1}^{n}\|f_i\|
\end{equation}
and $\omt (f)$ is essentially selfadjoint for all $f\in\K$.
\end{proposition}

\noindent
\textit{Proof.} Let $l(f)$ be the creation operator on the full Fock space 
over $\K$ and $j_n$ the restriction to $V_n$ of the isometry $j$. 
The main estimates are 
\begin{eqnarray}
\|a(f)\psi_k\|^2 
&=&\frac{1}{(k-1)!}
\|j^*_{k-1}\tens l^*(f)\psi_k\|_{V_{k-1}\tens\K^{\tens k-1}}^2 \leq\nonumber\\
&\leq & \frac{k!}{(k-1)!} \|f\|^2 \|\psi_k\|^2 = k\|f\|^2 \|\psi_k\|^2,
\end{eqnarray}
and similarly
\begin{equation}
\|a^*(f)\psi_k\|^2 \leq (k+1)\|f\|^2 \|\psi_k\|^2.
\end{equation}
This gives the same result as in the case of the symmetric Fock space
(Theorem X.41 in \cite{Reed.Simon.2}): 
\begin{equation}
 \|a^{\sharp}_{\mathbf{t}}(f_1)\dots a^{\sharp}_{\mathbf{t}}(f_n)\psi_k\|\leq 
\sqrt{(k+1)\dots (k+n)} \|\psi_k \| \prod_{i=1}^{n}\|f_i\|.  
\end{equation}
In particular the vectors with finite number of particles form a dense set $D$
of analytic vectors for the field operators $\omt (f)$. By Nelson's analytic 
vector theorem we conclude that $\omt (f)$ is essentially
selfadjoint.

\qed

\noindent
From now we will denote by the same symbol the closure of 
$\omt (f)$. We are now in the position to construct the von Neumann
algebras $\Gamma_{\mathbf{t}}(\K)$ as described in \ref{eq.gamma.k}
for any multiplicative positive definite $\mathbf{t}$. If 
$O:\K\ra \K'$ is an orthogonal operator between two Hilbert 
spaces we define its second quantisation as in
equation \ref{eq.sec.quant.orthogonal}.

\begin{corollary}\label{cor.expansion} 
Let $\psi_k\in \mathcal{F}_{\mathbf{t}}(\K)$ be a 
$k-$particles vector and $f_1,\dots ,f_n\in \K$. Then
\begin{equation}\label{eq.expansion}
\prod_{p=1}^{n}e^{i\omt (f_p)}\psi_k=\sum_{m_1,\dots ,m_n=0}^\infty 
\frac{(i\omt (f_1))^{m_1}\dots(i\omt (f_n))^{m_n}}{m_1!\dots m_n!}\psi_k.
\end{equation}
\end{corollary}

\noindent
\textit{Proof.} Using the previous proposition we get
\begin{eqnarray}
&&\sum_{m_1,\dots ,m_n=0}^\infty 
\frac{||\omt (f_1)^{m_1}\dots \omt (f_n)^{m_n}\psi_k||}
{m_1!\dots m_n!}\leq
\nonumber\\
&&\frac{||\psi_k||}{\sqrt{k!}}
\sum_{m_1,\dots ,m_p=0}^\infty
\frac{||f_1||^{m_1}\dots ||f_n||^{m_n} }{m_1!\dots m_n!}
\sqrt{(k+m_1+\dots +m_n)!}<\infty.\nonumber
\end{eqnarray}
This means that all vectors of the form $\prod_{p=1}^{n}e^{i\omt (f_p)}\psi_k$
are analytic for the field operators. In particular 
one can expand as in \ref{eq.expansion}. We denote the space of linear 
combinations of such ``exponential vectors'' by $D_e$. 
  
\qed

\begin{lemma}\label{lemma.isometry} 
Let $\K,\K'$ be real Hilbert spaces and $I:\K\ra\K'$ an isometry.
Then there exists an injective $^*$-homomorphism 
$\Gamma_{\mathbf{t}}(I) $ from $\Gamma_{\mathbf{t}}(\K)$ to 
$\Gamma_{\mathbf{t}}(\K')$.
\end{lemma}

\noindent
\textit{Proof.} There exists an orthogonal operator 
  $O_I:\K\oplus\ell^2(\Z)\ra\K'\oplus\ell^2(\Z)$ such that the restriction to 
$\K$ coincides with $I$. Then the map
\begin{equation}
\tilde{\Gamma}_{\mathbf{t}}(\K\oplus\ell^2(\Z))\ni X\mapsto 
\mathcal{F}_{\mathbf{t}}(O_I)X\mathcal{F}_{\mathbf{t}}(O_I^*) 
\in\tilde{\Gamma}_{\mathbf{t}}(\K'\oplus\ell^2(\Z))
\end{equation} 
sends an element $X\in\Gamma_{\mathbf{t}}(\K)$ into 
$\Gamma_{\mathbf{t}}(\K')$ when restricted to the subalgebra 
$\Gamma_{\mathbf{t}}(\K)$ and the restriction does not depend on the choice 
of the orthogonal $O_I$.

\qed

\begin{proposition} Let $P:\K\ra\K'$ be a coisometry i.e. 
$PP^*=\mathbf{1}_{\K'}$ and $\mathbf{t}$ a positive definite 
multiplicative function. Then 
\begin{equation}\label{eq.coisometry.alg}
\Gamma_{\mathbf{t}}(P):
X \mapsto \mathcal{F}_{\mathbf{t}}(P\oplus \mathbf{1}) X 
          \mathcal{F}_{\mathbf{t}}(P\oplus \mathbf{1})^*
\end{equation}
maps $\Gamma_{\mathbf{t}}(\K)$ onto $\Gamma_{\mathbf{t}}(\K')$.
\end{proposition}

\noindent
\textit{Proof.} We denote by $I$ the adjoint of $P$. 
We fix an orthonormal basis $(e_i)_{i=1}^\infty$ in 
$\K'\oplus\ell^2(\Z)$ and $(f_j)_{j=1}^M$ in  $\K\ominus(I\K')$. Let 
$X=\prod_{p=1}^{n}e^{i\lambda_p\omt (g_p)}$ be an element of
$\tilde{\Gamma}_{\mathbf{t}}(\K\oplus\ell^2(\Z))$ where each $g_p$ is either
an $(I\oplus\mathbf{1})e_i$ or an $f_j$. We will prove that 
$\mathcal{F}_{\mathbf{t}}(P\oplus \mathbf{1})
X\mathcal{F}_{\mathbf{t}}(P\oplus \mathbf{1})^* $ 
belongs to $\tilde{\Gamma}_{\mathbf{t}}(\K'\oplus\ell^2(\Z))$. Let 
\begin{equation}
Y=\tilde{\Gamma}_{\mathbf{t}}(O_{P})(X)=
\prod_{p=1}^{n}e^{i\lambda_p\omt (O_{P}g_p)}\in 
\tilde{\Gamma}_{\mathbf{t}}(\K'\oplus\ell^2(\Z))
\end{equation}
where $O_{P}:\K\oplus\ell^2(\Z)\ra\K'\oplus\ell^2(\Z)$ 
is an orthogonal operator which satisfies the condition that 
$O_Pg_p=(P\oplus\mathbf{1})g_p$ for all $g_p\in I\K'\oplus\ell^2(\Z)$. 
We denote by $\h_X$ the finite dimensional 
subspace of $\K'\oplus\ell^2(\Z)$ spanned by the vectors 
$(P\oplus\mathbf{1})g_p$ for 
$1\leq p\leq n$. Let $T$ be an operator which is of the form 
$\mathbf{1}_{\h_X}\oplus S$ with $S$ an arbitrary orthogonal operator which 
acts as a bilateral shift on the orthonormal basis of the orthogonal complement
of $\h_X$ in $\K'\oplus\ell^2(\Z)$. We claim that 
\begin{equation}
\wlim_{l\to \infty}\tilde{\Gamma}_{\mathbf{t}}(T^l)(Y)=
\mathcal{F}_{\mathbf{t}}(P\oplus \mathbf{1})
X\mathcal{F}_{\mathbf{t}}(P\oplus \mathbf{1})^*.
\end{equation}
It is sufficient to check this for expectation values with respect to 
vectors of the form $\psi(\mathcal{V},\mathbf{e})
=\Psi(\mathcal{V},\mathbf{e})\Omega_{\mathbf{t}}$  where
the components of $\mathbf{e}$ are elements of the basis $(e_i)_{i=1}^\infty$.
The linear span of such vectors forms the dense domain 
$D\subset\mathcal{F}_{\mathbf{t}}(\K'\oplus\ell^2(\Z))$. We apply now 
corollary \ref{cor.expansion} and find that for $l$ large enough:
\begin{eqnarray}
&&\br\psi(\mathcal{V},\mathbf{e}),
\tilde{\Gamma}_{\mathbf{t}}(T^l)(Y)
\psi(\mathcal{V},\mathbf{e})\ke =
\br\psi(\mathcal{V},~\mathbf{e}),
\prod_{p=1}^{n}e^{i\lambda_p\omt (T^lO_{P}g_p)}
\psi(\mathcal{V},\mathbf{e})\ke =
\nonumber\\
&&
\sum_{m_1,\dots ,m_n=0}^\infty
\br\psi(\mathcal{V},\mathbf{e}),~ 
\prod_{q=1}^n\frac{(i\lambda_q\omt (T^lO_{P}g_q))^{m_q}}{m_q!}
\psi(\mathcal{V},\mathbf{e})\ke = \nonumber\\
&& 
\sum_{m_1,\dots ,m_n=0}^\infty
\br\psi(\mathcal{V},I\mathbf{e}),~ 
\prod_{q=1}^n\frac{(i\lambda_q\omt (g_q))^{m_q}}{m_q!}
\psi(\mathcal{V},I\mathbf{e})\ke  =\nonumber\\
&=&
\br\psi(\mathcal{V},\mathbf{e}),~
\mathcal{F}_{\mathbf{t}}(P\oplus \mathbf{1})X
\mathcal{F}_{\mathbf{t}}(P\oplus \mathbf{1})^*\psi(\mathcal{V},\mathbf{e})\ke.
\nonumber
\end{eqnarray}
Indeed the pairing prescription of the fields of the same colour insures
that the terms in the two sums are equal one by one if we choose $l$ such that
no vector $T^lO_{P}g_p$ in the orthogonal complement of $\h_X$ coincides with
a component of $\mathbf{e}$. As the span of the operators of the form 
$\prod_{p=1}^{n}e^{i\lambda_p\omt (g_p)}$ is weakly dense in 
$\tilde{\Gamma}_{\mathbf{t}}(\K\oplus\ell^2(\Z))$ we can extend the map
\begin{equation}
\tilde{\Gamma}_{\mathbf{t}}(P\oplus\mathbf{1})(X)=
\mathcal{F}_{\mathbf{t}}(P\oplus\mathbf{1})X
\mathcal{F}_{\mathbf{t}}(P\oplus\mathbf{1})^*
\end{equation}
to the whole algebra such that 
$\tilde{\Gamma}_{\mathbf{t}}(P\oplus\mathbf{1})(X)\in
\tilde{\Gamma}_{\mathbf{t}}(\K'\oplus\ell^2(\Z))$. Now, if 
$X$ commutes with 
$\mathcal{F}_{\mathbf{t}}(1\oplus O)$ acting on 
$\mathcal{F}_{\mathbf{t}}(\K\oplus\ell^2(\Z))$
for $O\in\mathcal{O}(\ell^2(Z))$ then it is easy to see that
$\tilde{\Gamma}_{\mathbf{t}}(P\oplus\mathbf{1})(X)$ commutes with
$\mathcal{F}_{\mathbf{t}}(1\oplus O)$ acting on 
$\mathcal{F}_{\mathbf{t}}(\K'\oplus\ell^2(\Z))$. In other words the restriction
 $\Gamma_{\mathbf{t}}(P)$ of 
$\tilde{\Gamma}_{\mathbf{t}}(P\oplus\mathbf{1})$ to 
$\Gamma_{\mathbf{t}}(\K)$ has the desired property:
\begin{equation}
\Gamma_{\mathbf{t}}(P):\Gamma_{\mathbf{t}}(\K)\ra\Gamma_{\mathbf{t}}(\K').
\end{equation} 

\qed

\begin{corollary} Let $I:\K\ra\K'$ be an isometry. Then 
$\Gamma_{\mathbf{t}}(I^*)\Gamma_{\mathbf{t}}(I)=
\mathrm{id}_{\Gamma_{\mathbf{t}}(\K)}$. If $I':\K'\ra\K''$ is another isometry
then $\Gamma_{\mathbf{t}}(I')\Gamma_{\mathbf{t}}(I)=\Gamma_{\mathbf{t}}(I'I)$.

\end{corollary}

\noindent
\textit{Proof.} The map $\Gamma_{\mathbf{t}}(I^*)\Gamma_{\mathbf{t}}(I)$ is 
implemented by 
\begin{equation}
\Gamma_{\mathbf{t}}(I^*)\Gamma_{\mathbf{t}}(I):X\ra 
\mathcal{F}_{\mathbf{t}}((I^*\oplus\mathbf{1})O_I)X 
\mathcal{F}_{\mathbf{t}}((I^*\oplus\mathbf{1})O_I)^*.
\end{equation} 
But $(I^*\oplus\mathbf{1})O_I=\mathbf{1}\oplus Q $ where $Q$ is a coisometry
on $\ell ^2(\Z)$. Any such operator on 
$\ell^2(\Z)$ can be obtained as a weak limit of orthogonal operators. 
The functor $\mathcal{F}_{\mathbf{t}}$ respects weak limits 
and as $X$ commutes with all $\mathcal{F}_{\mathbf{t}}(\mathbf{1}\oplus O)$ 
for $O$ orthogonal operator, it also commutes with 
$\mathcal{F}_{\mathbf{t}}(\mathbf{1}\oplus Q)$, thus we get
\begin{equation}
\Gamma_{\mathbf{t}}(I^*)\Gamma_{\mathbf{t}}(I)(X)=
\mathcal{F}_{\mathbf{t}}(\mathbf{1}\oplus Q)X
\mathcal{F}_{\mathbf{t}}(\mathbf{1}\oplus Q)^*
=X.
\end{equation} 
The other identity follows directly from the definition of 
$\Gamma_{\mathbf{t}}(I)$.

\qed

\noindent
Any contraction $T:\K_1\rightarrow \K_2$ can be written as $T=PI$ where 
$I:\K_1\rightarrow \K$ is an isometry and 
$P:\K\rightarrow \K_2$ is a coisometry. This decomposition is not unique. We
define the second quantisation of $T$ by using the already constructed 
$\Gamma_{\mathbf{t}}(I)$ and $\Gamma_{\mathbf{t}}(P)$: 
\begin{equation}
\Gamma_{\mathbf{t}}(T):=\Gamma_{\mathbf{t}}(P) \Gamma_{\mathbf{t}}(I):
\Gamma_{\mathbf{t}}(\K_1)\ra\Gamma_{\mathbf{t}}(\K_2).
\end{equation} 
We will verify that $\Gamma_{\mathbf{t}}(T)$ does not depend on the choice of 
$I$ and $P$. Firstly we note that we can restrict only to ``minimal'' $\K$, 
that is, $\K$ is spanned by $I\K_1$ and $P^*\K_2$. If this is not the case 
then we make the decomposition $I=I_2I_1$ and $P=P_2I_2^*$
such that $T=P_2I_1$ is minimal and we use the previous corollary,
\begin{equation}
\Gamma_{\mathbf{t}}(T)=\Gamma_{\mathbf{t}}(P)\Gamma_{\mathbf{t}}(I)=
\Gamma_{\mathbf{t}}(P_2)\Gamma_{\mathbf{t}}(I_2^*) \Gamma_{\mathbf{t}}(I_2) \Gamma_{\mathbf{t}}(I_1)=\Gamma_{\mathbf{t}}(P_2) \Gamma_{\mathbf{t}}(I_1).
\end{equation}
Secondly, we compare two minimal decompositions $T=PI=P'I'$ with 
$I':K_1\ra\K'$. By minimality, there exists an orthogonal $O$ from $\K'$ 
and $\K$ defined by
\begin{eqnarray}
&&O:I'f\mapsto If,\qquad f\in\K_1\nonumber\\
&&O:P^{\prime *}g\mapsto P^*g,\qquad g\in\K_2\nonumber.
\end{eqnarray}
Then $PI=P'O^*OI'$ and by applying again the previous corollary we get 
$\Gamma_{\mathbf{t}}(P)\Gamma_{\mathbf{t}}(I)=\Gamma_{\mathbf{t}}(P')
\Gamma_{\mathbf{t}}(I')$.
\begin{lemma} For any contractions $T_1:\K_1\ra\K_2$ and $T_2:\K_2\ra\K_3$ we have
$\Gamma_{\mathbf{t}}(T_2)\Gamma_{\mathbf{t}}(T_1)=\Gamma_{\mathbf{t}}(T_2T_1)$.
\end{lemma}

\noindent
\textit{Proof.} The completely positive maps $\Gamma_{\mathbf{t}}(T_1)$ and 
$\Gamma_{\mathbf{t}}(T_2)$ are implemented by 
\begin{equation}
\Gamma_{\mathbf{t}}(T_i):
X\mapsto \mathcal{F}_{\mathbf{t}}(P_i)X\mathcal{F}_{\mathbf{t}}(P_i)^*
\end{equation}
with $P_i:\K_i\oplus\ell^2(\Z)\ra\K_{i+1}\oplus\ell^2(\Z)$ are coisometries 
with the matrix expression 
\begin{displaymath}
P_i=\left( \begin{array}{ccc}
T_i & A_i\\
0 & P_i'\\
\end{array}\right)
\end{displaymath}
for $i=1,2$. Their product $P_2P_1$ is a coisometry with matrix 
expression of the same form 
\begin{displaymath}
P_2P_1=\left( \begin{array}{ccc}
T_2T_1 & T_2A_1+A_2P_1'\\
 0     & P_2'P_1'\\
\end{array}\right).
\end{displaymath}
This implies that 
$\Gamma_{\mathbf{t}}(T_2T_1)=\Gamma_{\mathbf{t}}(T_2)\Gamma_{\mathbf{t}}(T_1)$.

\qed

By putting together all the results of this section we obtain the main theorem.
\begin{theorem} Let $\mathbf{t}$ be a positive definite function. Then 
there exists a functor of second quantisation $\Gamma_{\mathbf{t}}$ 
if and only if $\mathbf{t}$ is multiplicative and 
$\Gamma_{\mathbf{t}}(\emptyset)=\C$.
\end{theorem}

In the end we make the connection with the known cases of second quantisation.
 
\begin{corollary}\label{cor.faithfulvacuum}
Let $\mathbf{t}$ be a positive definite multiplicative function such that
the vector $\Omega_{\mathbf{t}}$ is cyclic and separating for 
$\tilde{\Gamma}_{\mathbf{t}}(\ell^2(\Z))$. Then we have the following:\\

\noindent 
1) the cyclic representation 
of $\Gamma_{\mathbf{t}}(\K)$ with respect to $\Omega_{\mathbf{t}}$
is faithful and the subspace of $\mathcal{F}_{\mathbf{t}}(\K\oplus\ell^2(\Z))$ 
spanned by $\Gamma_{\mathbf{t}}(\K)\Omega_{\mathbf{t}}$ is isomorphic to 
$\mathcal{F}_{\mathbf{t}}(\K)$. In this representation the second quantisation 
of a contraction $T:\K_1\ra\K_2$ is the 
completely positive map $\Gamma_{\mathbf{t}}(T)$ from 
$\Gamma_{\mathbf{t}}(\K_1)$ to $\Gamma_{\mathbf{t}}(\K_2)$ such that
\begin{equation}
\Gamma_{\mathbf{t}}(T)(X)\Omega_{\mathbf{t}}= 
\mathcal{F}_{\mathbf{t}}(T)X\Omega_{\mathbf{t}}
\end{equation}
for $X\in\Gamma_{\mathbf{t}}(\K_1)$. \\

\noindent 
2) if the field operators are 
bounded then $\Gamma_{\mathbf{t}}(\K)$ is the weak closure of the $^*$-algebra of generalised Wick products $\Psi_{\mathbf{t}}(\mathcal{V}, \mathbf{f})$ with all components $\mathbf{f}(i)\in\K\subset\K\oplus\ell^2(\Z)$.    
\end{corollary}

\noindent
\textit{Proof.} If $X\in\Gamma_{\mathbf{t}}(\K)$ then 
$\psi=X\Omega_{\mathbf{t}}$ is left invariant by 
$\mathcal{F}_{\mathbf{t}}(\mathbf{1}\oplus O)$ for all 
$O\in\mathcal{O}(\ell^2(\Z))$. This 
means that $\psi$ is orthogonal on all vectors of the form 
$\Psi(\mathcal{V}, \mathbf{e})\Omega_{\mathbf{t}}$ 
where $\mathbf{e}$ takes values in an orthogonal basis of 
$\K\oplus\ell^2(\Z)$ such that at least one of 
the components is an element of the basis in $\ell^2(\Z)$. By corollary 
\ref{cor.wick.cyclic} we conclude that the cyclic space of 
$\Gamma_{\mathbf{t}}(\K)$ is (up to a unitary isomorphism) 
$\mathcal{F}_{\mathbf{t}}(\K)\subset\mathcal{F}_{\mathbf{t}}
(\K\oplus\ell^2(\Z))$. According to the same corollary the generalised 
Wick products span the domain $D$ dense in $\mathcal{F}_{\mathbf{t}}(\K)$.

\section{An Example}\label{sec.applications}

In \cite{Boz.Sp.1} and \cite{Gu.Maa.} it
has been proved that for all $0\leq q \leq 1$, 
the following function on pair partitions is positive definite:
\begin{equation}\label{eq.t}
\mathbf{t}_q(\mathcal{V})=q^{|\mathcal{V}|-|\mathrm{B}(\mathcal{V})|}
\end{equation} 
where $|\mathrm{B}(\mathcal{V})|$ is the number of 
blocks of the pair partition $\mathcal{V}$. A block is a subpartition 
whose graphical representation is connected and does not intersect other  
pairs from the rest of the partition. The corresponding 
vacuum state 
$\rho_{\mathbf{t}_q}(\cdot)=
\br \Omega_{\mathbf{t}_q}, \cdot\Omega_{\mathbf{t}_q}\ke$ 
is tracial for any von Neumann algebra $\Gamma_{\mathbf{t}_q}(\K)$ 
associated to a real Hilbert space $\K$. Indeed for any pair partition
$\mathcal{V}$ we have
$\mathbf{t}_q(\mathcal{V})=
\br \Omega_{\mathbf{t}_q}, M_{\mathcal{V}}\Omega_{\mathbf{t}_q}\ke$ with 
$M_{\mathcal{V}}$ a monomial of fields containing $|\mathcal{V}|$ pairs of 
different colours arranged according to the pair partition $\mathcal{V}$. 
The trace property for the vacuum is equivalent with the 
invariance under circular permutations of the fields in the 
monomial $M_{\mathcal{V}}$ which is equivalent to the invariance of 
$\mathbf{t}_q$ under transformations described as follows:
\begin{eqnarray}
\mathcal{P}_2(\{1,\dots, 2r\})\ni \mathcal{V}  &\mapsto& 
\tilde{\mathcal{V}}\in\mathcal{P}_2(\{0,\dots, 2r-1\})  \\
\{p_1,\dots, p_{r-1}\}\cup \{ (l,2r)\}         &\mapsto& 
\{ (0,l)\}\cup\{p_1,\dots, p_{r-1}\}. 
\end{eqnarray}  
Under such transformations the number of blocks remains unchanged thus
$\mathbf{t}_{q}(\mathcal{V})$ is equal to 
$\mathbf{t}_{q}(\tilde{\mathcal{V}})$ and 
$\rho_{\mathbf{t}_q}$ is tracial. Thus the assumption of 
Corollary \ref{cor.faithfulvacuum} is satisfied and we have 
second quantisation at algebraic level.

\noindent
The version of $\mathbf{t}_q$ for $-1\leq q \leq 0$ is 
$\mathbf{t}_q:=\mathbf{t}_{-q}\mathbf{t}_{-1}$ where 
\begin{equation}
\mathbf{t}_{-1}(\mathcal{V})=(-1)^{|I(\mathcal{V})|}
\end{equation}
and $|I(\mathcal{V})|$ is the number of crossings of $\mathcal{V}$. 
The operators $\omega_{\mathbf{t}_q} (f)$ are bounded for $-1\leq q \leq 0$ 
\cite{Boz.Sp.1}. Thus by corollary \ref{cor.faithfulvacuum} the generalised 
Wick products form a strongly dense subalgebra of $\Gamma_{\mathbf{t}_q}(\K)$, faithfully represented on $\mathcal{F}_{\mathbf{t}}(\K)$.\\

\noindent
In the rest of this section we want to investigate the type of 
the von Neumann algebras $\Gamma_{\mathbf{t}_q}(\K)$ for 
$\mathrm{dim}~\K=\infty$ and $-1\leq q \leq 0$. Inspired by 
\cite{Boz.Ku.Spe.}, we will first find a sufficient condition for 
$\Gamma_{\mathbf{t}}(\K)$ to be a type $\mathrm{II}_1$ factor, 
and we will apply it to $\mathbf{t}_q$.

\noindent
Let $\mathbf{t}$ be a multiplicative positive definite function such that 
$\rho_{\mathbf{t}}$ is trace state on $\Gamma_{\mathbf{t}}(\K)$ for $\K$ 
infinite dimensional and such that $\omt (f)$ is bounded. 
Let $I$ be the natural isometry from $\K$ to 
$\K\oplus\R$, and $e_0$ a unit vector in the orthogonal complement 
of its image. The function $\mathbf{t}$ being multiplicative implies that
the map
\begin{equation}
\phi: \mathcal{F}_{\mathbf{t}}(\K)    \rightarrow 
      \mathcal{F}_{\mathbf{t}}(\K\oplus \R)
\end{equation}       
defined by $\phi=\omega_{\mathbf{t}}(e_0)\mathcal{F}_{\mathbf{t}}(I)$ 
is an isometry. 

\begin{definition}{\rm
Let $(P,L,R)$ be a disjoint partition of the ordered set \\
$\{1,\dots,2n+l+r\}$  and $d=(\mathcal{V},f_l,f_r)$ an  element of the 
$^*$-semigroup of broken pair partitions with 
$\mathcal{V}\in \mathcal{P}_2(P)$, $f_l:L\to \{1,\dots l\} $ the left legs and
 $f_r:R\to \{1,\dots r\}$ the right legs. We denote by 
$\underline{d}:=(\underline{\mathcal{V}},f_l,f_r)$ the element 
obtained by adding to $\mathcal{V}$ one pair  which embraces all other pairs
\begin{equation}
\underline{\mathcal{V}}:=\mathcal{V}\cup\{(0,2n+l+r+1)\}\in
\mathcal{P}_2(\{0\}\cup P\cup \{2n+l+r+1\}).
\end{equation}}
\end{definition}

\noindent
Then the map 
\begin{equation}
\Phi(\cdot):=
\Gamma_{\mathbf{t}}(I^*)(\omt (e_0)\Gamma_{\mathbf{t}}(I)(\cdot)\omt (e_0))
=\phi^*\Gamma_{\mathbf{t}}(I)(\cdot)\phi
\end{equation}
has the following action on the generalised Wick products:
\begin{equation}\label{eq.phiW}  
\Phi(\Psi(           \mathcal{V}, \mathbf{f}))=
             \Psi(\underline{\mathcal{V}},\mathbf{f})
\end{equation}
which on the level of von Neumann algebras gives the 
completely positive contraction from 
$\Gamma_{\mathbf{t}}(\K)$ to itself. 
We fix an orthonormal basis $(e_n)_{n=1}^\infty$ in $\K$. 
Then by direct computation one can check that:
\begin{equation} 
\Phi(X)=\mathop{\mathrm{w{-}lim}}_{n\to\infty}
\omega_\mathbf{t}(e_n)X \omega_\mathbf{t}(e_n). 
\end{equation}
Let now $\tau$ be an arbitrary tracial normal state on 
$\Gamma_{\mathbf{t}}(\K)$. Then using the fact that 
$\omega_\mathbf{t}(e_n)^2\ra\mathbf{1}$ weakly as $n\ra\infty$, we get:
\begin{equation}\label{eq.taudephi}  
\tau(\Phi(X))=\lim_{n\to\infty}
\tau(\omega_\mathbf{t}(e_n)X\omega_\mathbf{t}(e_n))=
\lim_{n\to\infty}\tau(\omega_\mathbf{t}(e_n)^2X)=\tau(X).
\end{equation}
Suppose now that 
\begin{equation} \label{eq.philak} 
\mathop{\mathrm{w{-}lim}}_{k\to\infty}\Phi^k(X)=
\rho_{\mathbf{t}}(X)\mathbf{1}
\end{equation}
 for all $X\in\Gamma_{\mathbf{t}}(\K)$ which by the faithfulness of the 
vacuum state is equivalent to 
$\lim_{k\to\infty}\Phi^k(X)\Omega_{\mathbf{t}}=
\rho_{\mathbf{t}}(X)\Omega_{\mathbf{t}}$. 
Then by equation (\ref{eq.taudephi}) we conclude that $\rho_{\mathbf{t}}$ 
is the only trace state on $\Gamma_{\mathbf{t}}(\K)$ which is thus a type 
$\mathrm{II}_1$ factor. Let us take a closer look at the contraction 
\begin{equation}\label{eq.C}
\Theta:X\Omega_{\mathbf{t}}\mapsto \Phi(X)\Omega_{\mathbf{t}}.
\end{equation}
From equation (\ref{eq.phiW}) the operator $\Theta$ commutes with the 
orthogonal projectors on the spaces with definite ``occupation numbers''
$\mathcal{F}_{\mathbf{t}}(n_1,\dots ,n_k)$ (see \ref{eq.occunumb}). Thus
\begin{equation}
\Theta :v\otimes_s e(\underline{n})\mapsto
\theta(v)\otimes_s e(\underline{n})
\end{equation}
where 
\begin{equation}
e(\underline{n}):=
\underbrace{e_1\tens\dots\tens e_1}_{n_1 \mathrm{times}}\tens
\dots\tens\underbrace{e_k\tens\dots\tens e_k}_{n_k \mathrm{times}},
\end{equation}
$v\in V_n$ and $\theta :V\rightarrow V$ is the linear operator defined by 
\begin{equation}\label{eq.c}
\theta :\chi_{\mathbf{t}}(d)\xi\mapsto\chi_{\mathbf{t}}(\underline{d})\xi,
\qquad (d\in\mathcal{B}\mathcal{P}_2(\infty)).
\end{equation}

\begin{lemma}
Let $\mathbf{t}$ be a multiplicative positive definite function such that 
$\rho_{\mathbf{t}}$ is trace. Then the operator 
$\theta :V\rightarrow V$ defined by \ref{eq.c} is a selfadjoint contraction.
\end{lemma}

\noindent\textit{Proof.} Let 
$d_1, d_2\in \mathcal{B} \mathcal{P}_2^{(n,0)}$ be two
diagrams with $n$ left legs and no right legs. Then
\begin{equation}
\br\chi_{\mathbf{t}}(d_1)\xi,~\theta~\chi_{\mathbf{t}}(d_2)\xi\ke_V=
\hat{\mathbf{t}}(d_1^*\cdot \underline{d_2}).
\end{equation}
But if $\rho_{\mathbf{t}}$ is a trace then  
\begin{equation}
\hat{\mathbf{t}}(d_1^*\cdot \underline{d_2})=
\hat{\mathbf{t}}(\underline{d_1}^*\cdot d_2)
\end{equation}
which implies that
\begin{equation}
\br\chi_{\mathbf{t}}(d_1)\xi,~\theta~\chi_{\mathbf{t}}(d_2)\xi\ke_V=
\br~ \theta~ \chi_{\mathbf{t}}(d_1)\xi,\chi_{\mathbf{t}}(d_2)\xi\ke_V.
\end{equation}
Thus $\theta$ is a selfadjoint contraction.

\qed

\begin{theorem} \label{th.type2}
If $\xi$ is the only eigenvector of $\theta$ with eigenvalue 1 then 
$\Gamma_{\mathbf{t}}(\K)$ is a 
$\mathrm{II}_1$ factor for any infinite dimensional real Hilbert space $\K$.
\end{theorem}

\noindent\textit{Proof.} 
The operator $\theta$ is a selfadjoint contraction, thus 
\begin{equation}
\mathop{\mathrm{w{-}lim}}_{k\to \infty}\theta^k=P_\xi
\end{equation} 
where $P_\xi$ is the projection on the subspace $\C\xi$.
This implies \ref{eq.philak} and thus $\Gamma_{\mathbf{t}}(\K)$ is 
a $\mathrm{II}_1$ factor .  

\qed

\begin{corollary} 
Let $\K$ be an infinite dimensional real Hilbert 
space and $\mathbf{t}_q$ the positive definte function for $-1< q\leq 0$. 
Then the von Neumann algebra $\Gamma_{\mathbf{t}_q}(\K)$ is 
a type $\mathrm{II}_1$ factor.
\end{corollary}

\noindent\textit{Proof.} 
Let $d_1, d_2\in \mathcal{B} \mathcal{P}_2^{(n,0)}$ be two
diagrams with $n\geq 1$ left legs and no right legs. Then
\begin{eqnarray}
 && 
\br\chi_{\mathbf{t}_q}(d_1)\xi, ~\theta^2 ~\chi_{\mathbf{t}_q}(d_2)\xi\ke_V=
  \hat{\mathbf{t}}_q(\underline{d_1}^*\cdot \underline{d_2})=
  q(-1)^n\cdot\hat{\mathbf{t}}_q(d_1^*\cdot \underline{d_2}) \nonumber\\
&&
 = q(-1)^n\cdot
\br\chi_{\mathbf{t}_q}(d_1)\xi,~ \theta~\chi_{\mathbf{t}_q}(d_2)\xi\ke_V.
\end{eqnarray} 
where we have used the selfadjointness of $\theta$ in the first step and 
\begin{eqnarray}
&& |B(\underline{d_1}^*\cdot \underline{d_2})|=|B(d_1^*\cdot \underline{d_2})|,
\nonumber\\
&&|\underline{d_1}^*\cdot \underline{d_2}|=|d_1^*\cdot \underline{d_2}|+1
\nonumber
\end{eqnarray}
in the second equality. Thus the restriction of $\theta$ 
to $V\ominus \C\xi$ has norm $|q|<1$ and we can apply Theorem \ref{th.type2}.

\qed

\noindent 
\textbf{Remark.} The case $0\leq q<1$ is technically more difficult as the 
field operators are unbounded. This will be treated separately in a 
forthcoming paper \cite{Gu.Maa.2}.\\

\noindent 
If $\rho_{\mathbf{t}}$ is a faithful, multiplicative, but non-tracial 
state for $\Gamma_{\mathbf{t}}(\K)$ then the operators 
$\phi,\Phi, \Theta, \theta$ can still be defined in the same way. 
If moreover, $\xi$ is the only eigenvector with eigenvalue 1 of 
the operator $\theta$, then by a similar argument 
it can be shown that the algebra $\Gamma_{\mathbf{t}}(\K)$ is a factor. 
Indeed if $X$ is an element in the center of $\Gamma_{\mathbf{t}}(\K)$ then 
$\Phi(X)= 
\mathop{\mathrm{w{-}lim}}_{k\to\infty}\omega_{\mathbf{t}}(e_n)^2X=X $. 
which contradicts the assumption on $\theta$. 
This factor cannot be of type $II_1$ because the vacuum state is not tracial.
Using this observation one can construct type $III$ factors for certain 
positive definite multiplicative functions on pair partitions. 
An example will be given in \cite{Gu.Maa.2}.

\vspace{2mm}

\noindent
\textit{Acknowledgements.} The authors would like to thank 
Marek Bo\.zejko and Roland Speicher for stimulating discussions and remarks.

\newpage

\vspace{2cm}

\vspace{2cm}

\newpage

\vspace{5cm}

\end{document}